\documentclass[12pt,a4]{article}

\usepackage{abstract}
\usepackage{amsfonts}
\usepackage{amsmath}
\usepackage{amssymb}
\usepackage{amsthm}
\usepackage{changepage}
\usepackage{color}
\usepackage[multiple]{footmisc}
\usepackage{graphicx}
\usepackage{multirow}
\usepackage[round]{natbib}
\usepackage{setspace}
\usepackage{subfig}
\usepackage{titlesec}

\newcommand{\estar}{e^\ast}
\newcommand{\tildeedoublestar}{\tilde{e}^{\ast\ast}}
\newcommand{\edoublestar}{e^{\ast\ast}}
\newcommand{\etriplestar}{e^{\ast\ast\ast}}

\newcommand{\tildeedoubledagger}{\tilde{e}^{\dagger\dagger}}
\newcommand{\edoubledagger}{e^{\dagger\dagger}}
\newcommand{\etripledagger}{e^{\dagger\dagger\dagger}}
\newcommand{\bartildeedoubledagger}{\bar{\tilde{e}}^{\dagger\dagger}}
\newcommand{\baretripledagger}{\bar{e}^{\dagger\dagger\dagger}}

\newcommand{\mbp}{\mathbf{p}}
\newcommand{\mbx}{\mathbf{x}}
\newcommand{\mby}{\mathbf{y}}

\newcommand{\mbpi}{\boldsymbol{\pi}}

\newcommand{\mbbR}{\mathbb{R}}

\renewcommand{\cal}{\mathcal}
\renewcommand{\epsilon}{\varepsilon}
\renewcommand{\geq}{\geqslant}
\renewcommand{\leq}{\leqslant}

\titleformat{\section}{\normalfont\scshape\centering}{\thesection}{1em}{}
\titleformat{\subsection}{\normalfont\em\centering}{\thesubsection}{1em}{}
\titleformat{\subsubsection}{\normalfont\em\centering}{\thesubsubsection}{1em}{}

\newtheorem{claim}{\normalfont\scshape Claim}

\allowdisplaybreaks

\begin{document}

\title{Ever since Ellsberg\thanks{The experiments reported in this paper were conducted in the Experimental Social Science Laboratory (Xlab) at UC Berkeley and the California Social Science Experimental Laboratory (CASSEL) at UCLA. This research has made use of the ALICE High Performance Computing Facility at the University of Leicester and BlueCrystal/BluePebble High Performance Computing at the University of Bristol. Financial support was provided by the National Science Foundation (Grant No. SES-0962543) and the British Academy/Leverhulme Trust (Grant No. SRG21\textbackslash 211614).}}

\date{\normalsize{\today}}

\author{Aluma Dembo, Shachar Kariv, Matthew Polisson, and John K.-H. Quah\thanks{Dembo: Reichman University (aluma.dembo@runi.ac.il); Kariv: University of California, Berkeley (kariv@berkeley.edu); Polisson: University of Leicester (matt.polisson@leicester.ac.uk); Quah: National University of Singapore (ecsqkhj@nus.edu.sg).}}

\maketitle

\begin{abstract} \linespread{1.1} \selectfont
Ellsberg's famous paradox challenged Savage's subjective expected utility theory (EUT)---which reduces uncertainty to risk---by suggesting an aversion toward ambiguity. We provide a revealed preference test of the full set of axioms underpinning subjective EUT under uncertainty and compare it to an analogous test of objective EUT under risk. We find that individual choices are as consistent with utility maximization and expected utility maximization under uncertainty as they are under risk. Nevertheless, there is greater empirical scope for non-EUT models under uncertainty than under risk, and the absolute and relative consistency of EUT and non-EUT models vary considerably across subjects.
\end{abstract}

\begin{abstract} \linespread{1.1} \selectfont
\textsc{JEL Codes:} D81, C91.
\end{abstract}

\begin{abstract} \linespread{1.1} \selectfont
\textsc{Keywords:} revealed preference, rationality, risk, uncertainty, ambiguity, subjective expected utility, probabilistic sophistication, multiple priors, $\alpha$-maxmin expected utility, experiment.
\end{abstract}

\newpage

\section{Introduction} \label{sec:intro}

The distinction between risk (known probabilities) and uncertainty (unknown probabilities) dates back at least to \cite{knight1921}. In \citeauthor{savage1954}'s (\citeyear{savage1954}) celebrated theory of subjective expected utility (EUT), however, the decision maker acts as if a single probability measure governs the states of the world, effectively reducing uncertainty to risk. \cite{ellsberg1961} famously countered this reduction using thought experiments suggesting an aversion toward ambiguity. Subsequently, a large experimental literature has documented Ellsberg-type behavior, and a large theoretical literature has developed models to accommodate it.

The objective of this paper is to provide a positive account of choice behavior under uncertainty---and to compare it to analogous behavior under risk---by evaluating the performance of subjective EUT as well as non-EUT models of ambiguity aversion. We do so in a choice environment where \emph{all} the axioms underpinning these models can be assessed jointly. Our tests are both \emph{comprehensive} and \emph{nonparametric}: comprehensive in that we evaluate complete preference representations rather than individual axioms, and nonparametric in that we rely only on revealed preference conditions, making no assumptions about the functional form of the underlying utility or the structure of the errors. Such tests are feasible only with rich individual-level data---analysis at the individual level, as opposed to pooling observations or assuming homogeneity across subjects, is crucial given the well-documented heterogeneity in attitudes toward ambiguity.

\paragraph{Data}

Our data come from the portfolio choice experiment of \cite{ahn2014}. A portfolio choice problem is a choice over state-contingent commodities: there are three states of nature, denoted $s = 1, 2, 3$, and for each state $s$ an Arrow security that pays one token (the experimental currency) if state $s$ is realized and nothing otherwise. An allocation therefore specifies a payoff in each state, so that choosing one amounts to dividing wealth across the three states. State 2 is assigned an objectively known probability $\pi_2 = \tfrac{1}{3}$, while states 1 and 3 have unknown probabilities $\pi_1 > 0$ and $\pi_3 > 0$ satisfying $\pi_1 + \pi_3 = \tfrac{2}{3}$. Letting $x_s \geq 0$ denote the demand for the security paying off in state $s$ and $p_s > 0$ its price, the budget constraint is $\mbp \cdot \mbx = 1$, where $\mbx = (x_1, x_2, x_3) \geq \mathbf{0}$ and $\mbp = (p_1, p_2, p_3) \gg \mathbf{0}$, and the subject may choose any portfolio satisfying the constraint.

Every such problem is a standard economic decision problem that can be interpreted either as a portfolio choice (the allocation of wealth among three assets) or as a consumer decision (the selection of a bundle of state-contingent commodities from a budget set). Each problem is presented through a user-friendly graphical interface, which allows the subject to make many choices from widely varying budget sets---generating the rich individual-level data on which our analysis depends. To isolate the role of uncertainty, we also draw on data from our earlier experiment \citep{dembo2026}, which is identical to the experiment of \cite{ahn2014} except that $\pi_1 = \pi_3 = \tfrac{1}{3}$ is objectively known and announced to subjects, eliminating any uncertainty.

\paragraph{Ellsberg Paradox}

A key feature of this decision problem is that subjects can limit their exposure to ambiguity by choosing allocations with payoffs that depend less on the uncertain states; in the limit, an allocation with $x_1 = x_3$ pays the same regardless of which uncertain state occurs, leaving the subject with no exposure to ambiguity whatsoever. To see how such choices can reveal attitudes toward ambiguity, consider the following version of the Ellsberg paradox in the portfolio choice setting, involving two budget sets. Suppose the allocation $\mbx^1$ (resp. $\mbx^2$) is chosen at the prices $\mbp^1$ (resp. $\mbp^2$) as follows:
\begin{center}
\begin{tabular}{cc|c}
  & $\mbp$ & $\mbx$ \\
\cline{2-3}
1 & \multicolumn{1}{|c|}{$(p_L, p_M, p_H)$} & $(x_M, x_H, x_L)$ \\
\hline
2 & \multicolumn{1}{|c|}{$(p_H, p_M, p_L)$} & $(x_L, x_H, x_M)$
\end{tabular}
\end{center}
where $p_H > p_M > p_L$ and $x_H > x_M > x_L$. The price of the security that pays off in the risky state (state 2) is the intermediate of the three prices, $p_M$, in both budgets, and its associated demand is also the highest, $x_H$, in each; the two securities that pay off in the uncertain states (states 1 and 3), by contrast, have their prices swapped between the first and second budgets, with their associated demands also swapping accordingly.

These choices are not compatible with the maximization of a utility function that is monotonic with respect to first-order stochastic dominance (FOSD) under any single prior $\mbpi = (\pi_1, \tfrac{1}{3}, \pi_3)$ satisfying $\pi_1 + \pi_3 = \tfrac{2}{3}$, and are therefore also incompatible with subjective EUT. In both budgets, the subject demands more of the security that pays off in the risky state than of the cheaper of the two uncertain securities, despite its higher price. In the first budget, where the cheaper uncertain security pays off in state 1, this is optimal under a single prior only if $\pi_1 < \tfrac{1}{3}$; in the second, where it pays off in state 3, it requires that $\pi_3 < \tfrac{1}{3}$. These cannot both hold, since $\pi_1 + \pi_3 = \tfrac{2}{3}$. Yet neither allocation is revealed preferred to the other, since neither was affordable at the prices prevailing when the other was chosen ($\mbp^1 \cdot \mbx^2 > \mbp^1 \cdot \mbx^1$ and $\mbp^2 \cdot \mbx^1 > \mbp^2 \cdot \mbx^2$), so these choices are consistent with utility maximization; in fact, under mild auxiliary assumptions, one could even choose the rationalizing utility function to belong to a class of multiple-prior models of ambiguity aversion, specifically \citeauthor{gilboa1989}'s (\citeyear{gilboa1989}) maxmin expected utility (see the Appendix for a proof).

\paragraph{Rationalizability Measures}

More broadly, revealed preference theory allows us to test the full hierarchy of \emph{rationalizability} conditions on a subject's entire dataset of many budget sets with widely varying prices. We do so by extending to uncertainty the nonparametric revealed preference approach to testing EUT against non-EUT alternatives under risk, developed by \cite{polisson2020} and employed by \cite{dembo2026}. For each subject, we first ask whether a dataset is \emph{rationalizable}, in the sense of being consistent with the maximization of a well-behaved (continuous and increasing) utility function. By \citeauthor{afriat1967}'s (\citeyear{afriat1967}) theorem, a dataset is rationalizable in this basic sense if and only if it obeys the generalized axiom of revealed preference (GARP). Extensions to \citeauthor{afriat1967}'s (\citeyear{afriat1967}) theorem then allow us to test whether a dataset is consistent with the maximization of utility functions that comply with a stronger set of axioms.

In the \cite{ahn2014} experiment, the probability of state 2 and the total probability of states 1 and 3 are commonly known to be $\tfrac{1}{3}$ and $\tfrac{2}{3}$, respectively. Furthermore, states 1 and 3 are treated symmetrically in the experiment, so there is no reason for subjects to believe that either uncertain state is more likely. Given these features of the experiment, it is then natural to require that the rationalizing utility function is (in addition to being well-behaved) symmetric in states 1 and 3 and increasing with respect to FOSD given the objectively known features of the underlying probability distribution. A dataset that can be rationalized by a utility function satisfying these properties is said to be \emph{FOSD-rationalizable under multiple priors} because these requirements are compatible with multiple-prior models that treat states 1 and 3 symmetrically.

A stronger hypothesis is that a subject is compatible with \emph{probabilistic sophistication} \citep{machina1992,epstein2000}, which requires that a dataset can be rationalized by a well-behaved utility function that is increasing with respect to FOSD for a fixed probability distribution over all three states. In the \cite{ahn2014} experiment, the symmetric treatment of states 1 and 3 means that the natural fixed distribution is the uniform prior $\mbpi = (\tfrac{1}{3}, \tfrac{1}{3}, \tfrac{1}{3})$. A dataset is said to be \emph{FOSD-rationalizable under the uniform prior} if it can be rationalized by a well-behaved utility function that is FOSD-increasing with respect to the uniform prior; this is equivalent to requiring that the rationalizing utility function is increasing and symmetric across all three states. Lastly, we can also check whether a dataset can be rationalized by a well-behaved utility function taking the expected utility form \citep{savage1954}, again with a uniform distribution over the states. A dataset is said to be \emph{EUT-rationalizable} if it is rationalizable in this sense.\footnote{It is possible to check whether a dataset can be rationalized by a utility function that is increasing with respect to FOSD under a \emph{non}-uniform distribution over the states (and the same is true for subjective EUT). We perform these tests and discuss the results in Section \ref{subsec:uniform}. It suffices to say at this point that the additional flexibility from non-uniform priors does not materially improve the performance of these models.}

Since GARP provides an exact test of basic rationalizability---either the data satisfy the axiom or they do not---and rich individual-level data almost always contain at least some violations, we assess how nearly the data comply with GARP using \citeauthor{afriat1973}'s (\citeyear{afriat1973}) critical cost-efficiency index (CCEI). The CCEI, denoted $\estar$ and bounded between 0 and 1, is the largest proportion of each budget constraint at which all violations of GARP are removed---the least costly adjustment that renders the data rationalizable. It can be interpreted as the decision maker leaving up to a fraction $1 - \estar$ of the budget on the table through inconsistent choices. The notion of CCEI can be extended to measure the proportion by which budget constraints must be reduced for the data to be FOSD-rationalizable or EUT-rationalizable, and recent advances in revealed preference analysis (see \cite{nishimura2017} and \cite{polisson2020}) allow for the calculation of these indices. For each subject, we thus report four CCEI-type scores: $\estar$ for (basic) rationalizability, $\tildeedoublestar \leq \estar$ for FOSD-rationalizability under multiple priors, $\edoublestar \leq \tildeedoublestar$ for FOSD-rationalizability under the uniform prior, and $\etriplestar \leq \edoublestar$ for EUT-rationalizability.

The nested CCEI-type scores let us decompose violations of subjective EUT and its non-EUT alternatives, comparing the magnitudes of these violations across the full set of axioms underpinning these models. Our test is thus both nonparametric and comprehensive: perfect consistency with a non-EUT model that obeys ordering and monotonicity with respect to FOSD under multiple priors---but not under the uniform prior and therefore not subjective EUT itself---would yield
\[
1 = \estar = \tildeedoublestar > \edoublestar \geq \etriplestar.
\]
Perfect consistency with subjective EUT, by contrast, would mean that all four rationalizability scores are equal to 1.

\paragraph{Summary of Results}

\begin{figure}[!t]
\begin{center}
\includegraphics[scale=.5]{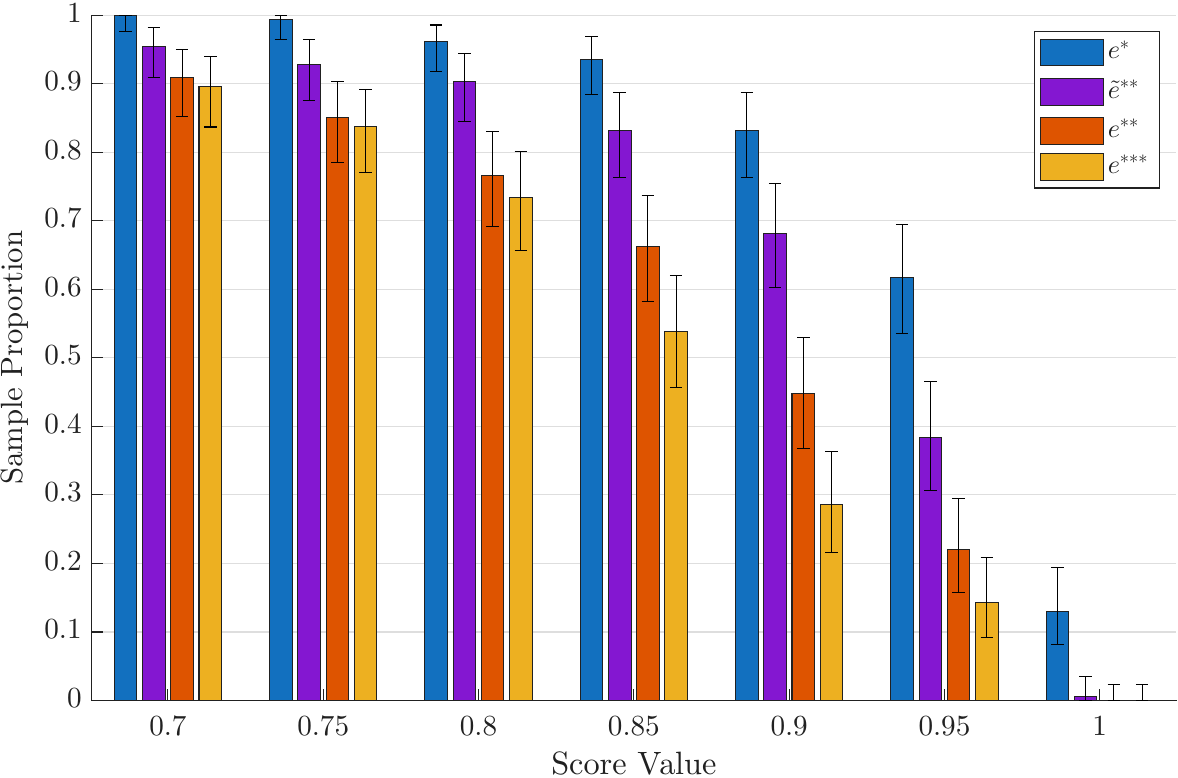}
\end{center}
\captionsetup{width=.95\linewidth}
\vspace{-.1in}
\caption{Distributions of Rationalizability Scores}
\vspace{-.1in}
\caption*{\footnotesize{The horizontal axis shows the score values and the vertical axis presents the fraction of subjects with scores above each value for rationalizability ($\estar$), FOSD-rationalizability ($\tildeedoublestar$, $\edoublestar$), and EUT-rationalizability ($\etriplestar$). The braces indicate 95 percent confidence intervals on these proportions.}}
\label{fig:figure1}
\end{figure}

Because the experiment provides a powerful test of revealed preference, very few subjects are perfectly rationalizable: out of 154 subjects only 20 (13.0 percent) are perfectly rationalizable ($\estar = 1$), one is perfectly FOSD-rationalizable under multiple priors ($\tildeedoublestar = 1$), and none are perfectly FOSD-rationalizable under the uniform prior ($\edoublestar = 1$)---and hence none are perfectly EUT-rationalizable ($\etriplestar = 1$). Figure \ref{fig:figure1} plots survival functions of the $\estar$, $\tildeedoublestar$, $\edoublestar$, and $\etriplestar$ rationalizability scores. For each score value on the horizontal axis, the vertical axis gives the fraction of subjects at or above that value.

Figure \ref{fig:figure1} reveals a nuanced pattern in which different gaps are more pronounced at different points of the score distribution. In the middle of the distribution---roughly the 0.8 to 0.9 range where the bulk of subjects are concentrated---the gap between $\tildeedoublestar$ and $\edoublestar$ is the most prominent, reflecting the drop in consistency from multiple-prior to uniform-prior FOSD-rationalizability, and pointing to some empirical scope for non-EUT models. Around the 0.9 level, however, a gap between $\edoublestar$ and $\etriplestar$ emerges, marking the further drop to EUT-rationalizability. At the upper tail, near 0.95 and above, both the gap between 1 and $\estar$ and the gap between $\estar$ and $\tildeedoublestar$ become prominent, meaning that at this level of consistency even basic rationalizability and multiple-prior FOSD-rationalizability are binding constraints. Taken together, the picture is one of layered complexity, with different rationalizability requirements binding for different subsets of subjects, which highlights the importance of our individual-level and comprehensive tests of all the axioms underpinning the different theories of choice under uncertainty.

We also compare our results to those obtained from the risk experiment in \cite{dembo2026}, which is identical to the \cite{ahn2014} experiment except that the probabilities of all three states are commonly known to be $\tfrac{1}{3}$. In the \cite{dembo2026} experiment, the $\estar$, $\edoublestar$, and $\etriplestar$ scores are also calculated for each subject ($\tildeedoublestar$ has no risk counterpart). Compared to the risk experiment, the distributions of rationalizability ($\estar$) and EUT-rationalizability ($\etriplestar$) scores under uncertainty are strikingly similar, implying that choices under uncertainty are as rationalizable and as EUT-rationalizable as choices under risk. The key difference is that under risk the gap between $\edoublestar$ and $\etriplestar$ is relatively modest for the vast majority of subjects, indicating little scope for the canonical non-EUT models that respect FOSD to outperform EUT. Under uncertainty, by contrast, the $\tildeedoublestar$ score introduces a new layer in the distribution, and the gap between $\tildeedoublestar$ and $\edoublestar$ is the dominant feature across much of the score range. This gap is precisely what creates the greater empirical scope for non-EUT models under uncertainty: while violations of ordering and monotonicity remain a primary source of departures from EUT in both domains, the additional wedge between multiple-prior and uniform-prior FOSD-rationalizability under uncertainty opens room for non-EUT models that find little scope under risk.

There are a variety of multiple-prior models of attitudes toward ambiguity, though, as \cite{ahn2014} shows, they mainly give rise to one of two specifications. The first is a ``kinked'' specification, which can be derived as a special case of a variety of utility models in the literature, including maxmin expected utility, Choquet expected utility, $\alpha$-maxmin expected utility ($\alpha$-MEU), and contraction expected utility. The second is a ``smooth'' specification, based on the class of recursive expected utility models.\footnote{The specifications are so called because of the shape of the indifference curves between the two uncertain securities: in the kinked specification, except for the special case of subjective EUT, the indifference curves have a kink along the ambiguity-free diagonal $x_1 = x_3$, so such allocations are optimal for a non-negligible range of prices, while the smooth specification is everywhere differentiable, so $x_1 = x_3$ is chosen only for a negligible set of prices.}$^{\text{,}}$\footnote{For the kinked specification, see maxmin expected utility: \cite{gilboa1989}; Choquet expected utility: \cite{schmeidler1989}; $\alpha$-MEU: \cite{ghirardato2004} and \cite{olszewski2007}; and contraction expected utility: \cite{gajdos2008}. For the smooth specification, see \cite{klibanoff2005}, \cite{nau2006}, \cite{ergin2009}, \cite{seo2009}, and also \cite{giraud2005}, \cite{halevy2005}, and \cite{ahn2008}, among others.} The standard subjective EUT model is a special case of both the kinked and smooth specifications. Because both specifications are accommodated by FOSD-rationalizability under multiple priors, their empirical performance is bounded above by $\tildeedoublestar$ (and below by $\etriplestar$).

In the final piece of analysis, we move from model-free bounds to a direct test of a specific model. We calculate a CCEI-type score for the kinked specification, denoted by $e^{\kappa}$ ($\kappa$ for kinked), interpreting the specification through the lens of $\alpha$-MEU.\footnote{The difficulty in testing the smooth specification stems largely from its dependence on the cardinal scale of utility, as pointed out by \cite{ahn2014}: its ambiguity and scale parameters enter only as a product and so are not separately identified absent a normalization. This cardinal indeterminacy, which \cite{ahn2014} flags as a parametric identification problem, reappears at the revealed preference level as a nonparametric identification problem. We elaborate on this issue in Sections \ref{sec:models} and \ref{sec:analysis}.} The test is nonparametric in the same sense as our other tests: for each value of the ambiguity parameter $\alpha$, it fixes only the weights that the kinked specification places on the worst- and best-case uncertain payoffs and leaves the Bernoulli index unrestricted. Since $\etriplestar \leq e^{\kappa} \leq \tildeedoublestar$, the score splits the multiple-prior scope $\tildeedoublestar - \etriplestar$ into the part that the kinked specification captures and a residual that bounds what any other multiple-prior model, including the smooth specification, could potentially add.

We find that the kinked specification and probabilistic sophistication are complementary rather than competing explanations. Because $e^{\kappa}$ is not bounded above or below by $\edoublestar$, the two scores can be compared directly, and the comparison is systematic: across the middle of the score distribution the kinked specification accommodates more subjects than probabilistic sophistication (83.1 versus 76.6 percent of subjects have scores above 0.8), and at moderate levels of consistency it accommodates nearly as many as the entire multiple-prior class (95.5 percent above 0.7 under both), whereas at the highest levels of consistency the ordering reverses (44.8 versus 39.6 percent above 0.9 in favor of probabilistic sophistication). The best-fitting ambiguity parameter is tightly concentrated just above the subjective EUT benchmark $\alpha = \tfrac{1}{2}$, consistent with mild ambiguity aversion on average. At the individual level, the sample splits almost evenly between subjects whose choices are better explained by the kinked specification ($\edoublestar < e^{\kappa}$, 42.9 percent) and subjects whose choices are better explained by probabilistic sophistication ($\edoublestar > e^{\kappa}$, 44.8 percent).

Taken together, the better of the two models for each subject, $\max \, \{e^{\kappa}, \edoublestar\}$, recovers much of the multiple-prior scope $\tildeedoublestar - \etriplestar$: the fraction of subjects with scores above 0.9 rises from 28.6 percent under subjective EUT to 50.0 percent, against 68.2 percent for $\tildeedoublestar$. The improvement over subjective EUT, $\max \, \{e^{\kappa}, \edoublestar\} - \etriplestar$, averages 0.038 across subjects (0.027 at the median), and given the money-metric interpretation of the CCEI, it reduces the fraction of the budget left on the table by more than 2.5 percent of wealth for over half of the subjects and by more than 5 percent for more than a quarter. These results show that the empirical wedge introduced through $\tildeedoublestar$ is not merely room that some unspecified model could in principle fill: a prominent and parsimonious model of ambiguity aversion, with a single parameter beyond the Bernoulli index, together with probabilistic sophistication, fills much of it. At the same time, the heterogeneity across subjects cautions against adopting any single model, EUT or non-EUT, as a universal description of choice under uncertainty.

\paragraph{Outline}

The rest of the paper is organized as follows. Section \ref{sec:lit} provides some background and situates our contribution within the wider literature. Section \ref{sec:data} introduces the experimental data setting, and Section \ref{sec:models} describes the main models of choice under uncertainty. Section \ref{sec:analysis} outlines our revealed preference analytical approach. Section \ref{sec:subjects} discusses some illustrative subjects to highlight basic features of the individual-level data, and Section \ref{sec:results} presents the empirical results. Section \ref{sec:conclusion} concludes.

\section{Background and Motivation} \label{sec:lit}

We do not attempt to review the large and growing theoretical-experimental literature on ambiguity aversion. Instead, we focus on the papers most relevant to our study. \cite{etner2012}, \cite{wakker2012}, and \cite{machina2014} provide comprehensive treatments of both theory and experimental evidence, while \cite{trautmann2015} provides a recent survey focused primarily on the experimental literature, and \cite{camerer1992} and \cite{camerer1995} offer excellent, if now somewhat dated, earlier surveys. Key experimental papers include \cite{halevy2007}, \cite{bossaerts2010}, \cite{hey2010}, \cite{conte2013}, \cite{ahn2014}, \cite{baillon2015}, \cite{dimmock2016}, and \cite{baillon2018}.

\cite{halevy2007} develops an experiment specifically designed to examine the connection between the reduction of objective compound lotteries and attitudes toward ambiguity. Four different urns are used to elicit choices in the presence of risk, uncertainty, and two degrees of compound uncertainty. EUT and non-EUT models of choice under uncertainty, as well as under both risk and uncertainty, generate different predictions about how reservation values for these four urns will be ordered, allowing each subject to be classified according to which model best predicts their ordering. \cite{halevy2007} concludes that no single model accounts for all observed behaviors---all models are represented in the subject pool.

\cite{ahn2014} shares \citeauthor{halevy2007}'s (\citeyear{halevy2007}) starting point that different models may be appropriate for describing different subjects' behavior, but takes this view of heterogeneity further: not only is there variation in the appropriateness of different models across subjects, but also variation in the degree of ambiguity aversion among subjects who conform to the same model. Specifically, the experiment of \cite{ahn2014} enables estimation of parametric models at the level of the individual subject: a kinked specification, a smooth specification, and a ``generalized kinked'' specification. The first two are each characterized by two parameters, one measuring risk aversion and the other ambiguity aversion, while the generalized kinked specification adds a third parameter for pessimism/optimism and can be interpreted as a recursive non-EUT model in the sense of \cite{segal1987,segal1990}. As shown in Section \ref{sec:models}, choices consistent with any of these specifications are FOSD-rationalizable under multiple priors but not necessarily under a single prior. The parameter estimates exhibit considerable heterogeneity across all specifications, but subjective EUT cannot be rejected for a majority of subjects, while most of the remaining subjects exhibit statistically significant ambiguity aversion (or seeking) and/or pessimism (or optimism).

A limitation shared by \cite{halevy2007}, \cite{ahn2014}, and other work measuring ambiguity aversion, is the potential for model misclassification. Assigning a model to each subject may conflate ambiguity attitudes with failures of the more basic axioms that underpin all of these models. This is true specifically for subjects who are not FOSD-rationalizable under multiple priors and/or not even rationalizable in the sense of ordering (completeness and transitivity). \cite{halevy2018} makes the analogous point in the domain of risk: structural estimation cannot separate risk preferences from model misspecification and choice inconsistency, so that failures of basic rationalizability are absorbed into the estimated parameters rather than detected. The nonparametric revealed preference approach we adopt confronts precisely this difficulty under uncertainty---as \cite{dembo2026} shows under risk---isolating failures of rationalizability and FOSD-rationalizability before any residual departure from subjective EUT is attributed to ambiguity attitudes.

Concretely, using the data of \cite{ahn2014}, we extend the revealed preference framework of \cite{dembo2026} from risk to uncertainty, providing the first comprehensive, nonparametric, individual-level test of the full set of axioms underlying subjective EUT and prominent non-EUT alternatives. Just as \cite{dembo2026} shows that violations of EUT under risk run much deeper than violations of independence alone---with departures from ordering and monotonicity dwarfing departures from the independence axiom---we ask whether the same is true in an otherwise identical setting under uncertainty, developing the analogous tests. We thus decompose violations of subjective EUT and its non-EUT alternatives in a way that no prior study has attempted, offering new guidance for theorists building models of ambiguity aversion and for experimentalists designing tests of such models.

\section{Experimental Data} \label{sec:data}

The experimental data we analyze were originally collected by \cite{ahn2014} to estimate parametric models of ambiguity aversion at the individual level. The key feature of the design is that it allows subjects to make numerous choices over a wide range of budget sets, yielding rich datasets well-suited to individual-level analysis without the need to pool data or assume homogeneity across subjects. Because choices are from standard budget sets, we are able to use revealed preference analysis to assess the consistency of the individual-level data with the entire set of axioms underlying the canonical models of choice under uncertainty.

In the experiment, \cite{ahn2014} presented subjects with a sequence of portfolio choice problems involving three assets---the selection of bundles of contingent commodities from three-dimensional budget sets. There are three states of nature denoted by $s = 1, 2, 3$, and for each state $s$, an Arrow security that pays one token (the experimental currency) in state $s$ and nothing in the other states. To distinguish between the effects of risk and uncertainty, state 2 has an objectively known probability, whereas the probabilities of states 1 and 3 are unknown. Specifically, state 2 occurs with probability $\pi_2 = \tfrac{1}{3}$ and states 1 and 3 occur with unknown probabilities $\pi_1 > 0$ and $\pi_3 > 0$. Subjects were only informed that $\pi_1 + \pi_3 = \tfrac{2}{3}$. Except for the presence of uncertainty, the experiment of \cite{ahn2014} is otherwise identical to the risk experiment of \cite{dembo2026}, where all states are equally likely: $\pi_1 = \pi_2 = \pi_3 = \frac{1}{3}$. We use the data of \cite{dembo2026} for comparison in our analysis below.

Letting $x_s \geq 0$ denote the demand for the security paying off in state $s$ and $p_s > 0$ denote its corresponding price, and (without loss of generality) normalizing the experimental budget to 1, the budget set can be written as
\[
\cal{B} = \{\mbx \in \mbbR^3_+ : \mbp \cdot \mbx \leq 1 \},
\]
where $\mbx = (x_1, x_2, x_3)$ and $\mbp = (p_1, p_2, p_3)$. Decision problems were presented using the graphical interface developed by \cite{choi2007b} and first used in \cite{choi2007a}, through which subjects could choose any allocation $\mbx \geq \mathbf{0}$ on the budget constraint $\mbp \cdot \mbx = 1$.\footnote{\cite{choi2014} uses the same graphical methodology to study risk preferences in a nationally representative sample from the Netherlands, and \cite{cappelen2023} in student samples from the US and Tanzania. Three-dimensional budget sets were first used by \cite{fisman2007} to study social preferences. A series of subsequent papers have applied similar methodologies to different pools of subjects \citep{fisman2015a,fisman2017,fisman2015b,fisman2023,li2017,li2022}. \cite{zame2026} combines the risk- and social-preference designs, eliciting choices across personal-risk, social-choice, and social-risk domains from the same subjects to link risk and social preferences.}

The experiment consisted of 50 decision rounds. In each round, subjects allocated tokens between three token accounts ($X$, $Y$, and $Z$), each corresponding to an axis in a three-dimensional graph scaled from 0 to 100 tokens. Which account $X$, $Y$, or $Z$ paid off in the risky state was rotated across experimental sessions, counterbalancing any framing or position effects tied to a particular account. Each choice involved selecting a point on a budget set of feasible token allocations. For each round, the computer randomly selected a budget set from those intersecting at least one axis at 50 or more tokens, with no intercept below 10 tokens or above 100 tokens. The budget sets selected for each subject in his decision problems were independent of each other and of the budget sets selected for other subjects in their decision problems. Subjects were not informed of which account was actually selected at the end of each round. At the end of the experiment, one decision round was randomly selected for payment, with each round having an equal probability of being chosen.\footnote{The random payoff method---paying one randomly selected decision---has been much debated. A subject who integrates decisions together with the randomizing device, rather than evaluating each in isolation, can use the objective randomization to hedge ambiguity \citep{oechssler2014,bade2015,saito2015,kuzmics2017,azrieli2018,ke2020}. In an Ellsberg design, \cite{baillon2022b} finds that the random incentive system roughly halves measured ambiguity aversion, with about half of ambiguity-averse subjects integrating. These concerns bear much less directly on the \cite{ahn2014} data analyzed here: subjects make 50 choices over heterogeneous budget sets, presented sequentially and with only the random price-generating process disclosed, so the cross-decision coordination needed to hedge is largely unavailable. \cite{ahn2014} makes this point, as does \cite{baillon2022a}, placing the portfolio design among the tasks least susceptible to random-incentive hedging.}

Let $\cal{D} = \{(\mbp^i, \mbx^i)\}_{i = 1}^{50}$ be the data generated by a given subject's choices, where $\mbp^i$ denotes the $i$-th observation of the price vector and $\mbx^i$ denotes the corresponding allocation. We combine two datasets: \cite{ahn2014}, with 154 subjects making decisions under uncertainty, and \cite{dembo2026}, with 168 subjects making decisions under risk. We emphasize that the subjects of \cite{ahn2014} and \cite{dembo2026} made choices over three-dimensional budget sets. The subjects of \cite{choi2007a} and \cite{choi2014} made choices over two-dimensional budget lines, and these datasets have been analyzed by \cite{halevy2018}, \cite{polisson2020}, \cite{declippel2023}, and \cite{echenique2023}, among others. The three-dimensional design has two important advantages over the two-dimensional one: with three goods, the weak and generalized axioms of revealed preference---WARP and GARP---are no longer observationally equivalent, so the design can distinguish them and thereby separate incompleteness from nontransitivity; and it provides a stronger test of the revealed preference conditions. We discuss this in greater detail in Section \ref{sec:analysis}; see also \cite{dembo2026}.

\section{Models of Choice under Uncertainty} \label{sec:models}

\subsection{Single-Prior Models} \label{subsec:single}

In our experimental context, standard models of choice under uncertainty all specify an increasing utility function $U : \mbbR^3_+ \to \mbbR$, with the precise form of $U$ varying to accommodate different behavioral regularities and/or anomalies. The canonical model of choice under uncertainty is subjective EUT \citep{savage1954}, where utility is evaluated as the expectation of an increasing Bernoulli index under a single (unique) prior. Because in the \cite{ahn2014} experiment the probability of state 2 is commonly known to be $\tfrac{1}{3}$ and states 1 and 3 are treated symmetrically, it is natural that the utility function also treats states 1 and 3 symmetrically, implying that the most natural single prior is the uniform prior, under which subjective EUT takes the following form:
\[
U(\mbx) = \frac{1}{3} \, u(x_1) + \frac{1}{3} \, u(x_2) + \frac{1}{3} \, u(x_3),
\]
where $u : \mbbR_+ \to \mbbR$ is the Bernoulli index. It is well known in general that subjective EUT is FOSD-increasing under a single prior, or that whenever $\mbx$ FOSD-dominates $\mbx'$, the former gives higher utility. In particular, this holds for subjective EUT under the uniform prior.

A generalization of subjective EUT involves \emph{probabilistic sophistication} \citep{machina1992,epstein2000}, which simply requires that utility is FOSD-increasing under a single prior; the model retains the single-prior feature of subjective EUT but dispenses with its specific functional form. In our experimental context, the only single prior consistent with the exogenously given probability of state 2 and the symmetry between states 1 and 3 is again the uniform prior. It is straightforward to check that a utility function is FOSD-increasing under the uniform prior if and only if it is increasing and symmetric across all three states: $U(\mbx) = U(\mbx')$ whenever the entries of $\mbx'$ are a permutation of the entries in $\mbx$.

\subsection{Multiple-Prior Models} \label{subsec:multiple}

Probabilistic sophistication cannot account for the Ellsberg-type behavior presented in the Introduction because it does not allow for ambiguity-sensitive behavior. Models that allow for such behavior typically introduce multiple priors over the states, which are then incorporated into utility in different ways. As noted in Section \ref{sec:lit}, many multiple-prior models give rise to one of two main utility specifications---kinked and smooth---and subjective EUT is nested within both.\footnote{These specifications also appear in a more restrictive form in \cite{ahn2014}, where they are estimated parametrically. In \cite{ahn2014}, the Bernoulli index takes a constant absolute risk aversion (CARA) form throughout and the second-order distribution in the smooth specification is assumed to be uniform; in our analysis, we leave the Bernoulli index unrestricted beyond continuity and monotonicity, and require only that the second-order distribution is symmetric.} We now outline the kinked and smooth specifications, focusing on their functional forms as a preface to establishing, at the end of this section, that both are FOSD-increasing under multiple priors. Any behavior compatible with these specifications is therefore FOSD-rationalizable under multiple priors but not necessarily under a single prior.

\paragraph{Kinked Specification}

In our experimental context, the kinked specification takes the following functional form:
\begin{equation}
U(\mbx) = \frac{2}{3} \, \alpha \, u(x_{\min}) + \frac{1}{3} \, u(x_2) + \frac{2}{3} \, (1 - \alpha) \, u(x_{\max}), \label{eq:kinked}
\end{equation}
where $x_{\min} = \min \, \{x_1, x_3\}$ and $x_{\max} = \max \, \{x_1, x_3\}$, and $\alpha \in (0, 1)$ is the ambiguity parameter. Utility is thus a weighted average of expected utility in the worst- and best-case scenarios across the two uncertain states. This specification is called ``kinked'' because for any $\alpha \neq \tfrac{1}{2}$ the indifference curves have a kink along the $x_1 = x_3$ diagonal, where the allocations are ambiguity-free. If $\alpha = \tfrac{1}{2}$ the kinked specification reduces to standard subjective EUT; higher values of $\alpha$ lead to greater ambiguity aversion, with $\alpha > \tfrac{1}{2}$ (resp. $\alpha < \tfrac{1}{2}$) indicating ambiguity aversion (resp. seeking) (see the Appendix for proofs of these claims).

In the $\alpha$-MEU model \citep{ghirardato2004}, the utility function has the form $U(\mbx) = \bar{\alpha} \min_{\mbpi \in \Pi} \mathbb{E}_{\mbpi} u(\mbx) + (1 - \bar{\alpha}) \max_{\mbpi \in \Pi} \mathbb{E}_{\mbpi} u(\mbx)$, where $\mathbb{E}_{\mbpi} u(\mbx)$ is the expected utility of $\mbx$ under the prior $\mbpi$, $\Pi$ is a closed and convex set of priors, and $\bar{\alpha} \in [0,1]$. It is clear that the kinked specification is a special case of this model: if we assume that $\Pi$ is the entire set of distributions consistent with the objective information in the experiment ($\Pi = \{\mbpi : \pi_2 = \tfrac{1}{3}\}$), then we obtain the formula in equation (\ref{eq:kinked}) if we set $\bar{\alpha} = \alpha$, since the minimum (resp. maximum) over $\Pi$ is attained at the prior that places the entire weight $\tfrac{2}{3}$ on the uncertain state with the lower (resp. higher) payoff.

However, even more can be said: the $\alpha$-MEU model leads to the kinked specification for any $\bar{\alpha} \in [0,1]$ so long as $\Pi$ is symmetric in states 1 and 3, in the sense that if $\mbpi = (b, \tfrac{1}{3}, c)$ is in $\Pi$ then so is $\mbpi' = (c, \tfrac{1}{3}, b)$, which leads to $\Pi = \{(a_1, \tfrac{1}{3}, a_3) : a_1 + a_3 = \tfrac{2}{3} \; \text{and} \; a_1, a_3 \geq \underline{a}\}$ for some $\underline{a} \in (0, \tfrac{1}{3}]$ (see the Appendix for details).\footnote{In particular, this means that the weight $\bar{\alpha}$ and the set of priors $\Pi$ in the $\alpha$-MEU model cannot be separately identified. For example, suppose that $U(\mbx) = \tfrac{1}{2} u(x_{\min}) + \tfrac{1}{3} u(x_2) + \frac{1}{6} u(x_{\max})$. This can be explained by setting $\bar{\alpha} = \tfrac{3}{4}$ and $\Pi = \{\mbpi : \pi_2 = \tfrac{1}{3}\}$, but it can also be explained by setting $\bar{\alpha} = 1$ and $\Pi = \{(a, \tfrac{1}{3}, \tfrac{2}{3} - a) : \tfrac{1}{6} \leq a \leq \tfrac{1}{2}\}$. This indeterminacy issue with the $\alpha$-MEU model is known (see \cite{siniscalchi2006}) and is not peculiar to our context. See also the discussion in \cite{ahn2014}.} Apart from the $\alpha$-MEU model, the kinked specification also arises from other models of choice under uncertainty, such as the Choquet expected utility with a convex capacity \citep{schmeidler1989} and the contraction model with fixed objective information \citep{gajdos2008} (see Appendix V of \cite{ahn2014}).

\paragraph{Smooth Specification}

In contrast to the kinked specification, the smooth specification is everywhere differentiable (hence its name), so that increased ambiguity aversion affects behavior in a manner qualitatively similar to increased risk aversion. The smooth model of choice under uncertainty \citep{klibanoff2005} leads, in our experimental context, to the following recursive double-expectation functional form:
\begin{equation}
U(\mbx) = \int_{-\frac{1}{3}}^{\frac{1}{3}} \varphi \left ( \left( \frac{1}{3} - \epsilon \right) \, u(x_1) + \frac{1}{3} \, u(x_2) + \left( \frac{1}{3} + \epsilon \right) u(x_3) \right) \mu(\epsilon) \, d\epsilon, \label{eq:smooth}
\end{equation}
where $\varphi : u(\mbbR_+) \to \mbbR$ is an increasing, possibly nonlinear transformation over expected utility levels and $\mu : (-\tfrac{1}{3}, \tfrac{1}{3}) \to \mbbR_+$ is a (second-order) density over possible priors, keeping the probability of state 2 fixed at $\tfrac{1}{3}$. Given the symmetric treatment of states 1 and 3, it is natural to require that $\mu$ is symmetric about $\epsilon = 0$. The concavity (resp. convexity) of the aggregator $\varphi$ determines the degree of ambiguity aversion (resp. seeking), with linear $\varphi$ reducing the smooth specification to standard subjective EUT.

We can interpret this model in the following way: unsure which of the possible first-order prior beliefs actually governs the states, the decision maker evaluates expected utility at each prior, transforms them by the function $\varphi$, before integrating these utilities with respect to the second-order distribution $\mu$. In this model, the restrictions on choice depend jointly on $u$, $\varphi$, and $\mu$, none of which are observed; because $\varphi$ acts on expected utility levels, the restrictions remain nonlinear in $u$ even for fixed $\varphi$ and $\mu$, in contrast to the kinked specification, where fixing $\alpha$ leaves restrictions that are linear in $u$. This makes the smooth specification difficult to test nonparametrically, and we do not attempt to do so; see also Section \ref{subsubsec:kinked-rat}.

\vspace{.25in}

Common to both the kinked and smooth specifications in equations (\ref{eq:kinked}) and (\ref{eq:smooth}) are two important structural properties: ($i$) utility is symmetric across the two uncertain states ($U(a, b, c) = U(c, b, a)$), and ($ii$) utility is FOSD-increasing under the objectively known probability information (state 2 occurs with probability $\tfrac{1}{3}$, while states 1 and 3 occur with probability $\tfrac{2}{3}$). For a well-behaved utility function, property ($ii$) is equivalent to requiring that $U(a, b, a) > U(b, a, b)$ whenever $a > b$ (since the probability of the higher outcome $a$ is greater in $(a, b, a)$ compared to $(b, a, b)$). Since choosing FOSD-dominated allocations is generally considered to be a mistake, all utility models of choice under risk and uncertainty (including but not restricted to the models behind the kinked and smooth specifications in equations (\ref{eq:kinked}) and (\ref{eq:smooth})) should satisfy property ($ii$). We refer to any well-behaved utility function satisfying both properties ($i$) and ($ii$) as \emph{FOSD-increasing under multiple priors}. Any such utility function allows a decision maker to behave as though the division of the probability between states 1 and 3 varies from choice to choice; it accommodates both ambiguity-averse and ambiguity-seeking behavior. Naturally, any utility function that is FOSD-increasing under the uniform prior is also FOSD-increasing under multiple priors.\footnote{Beyond the kinked and smooth specifications, \cite{ahn2014} also considers a generalized kinked specification---a kinked specification with an additional pessimism/optimism parameter---which can be interpreted as a recursive non-expected utility model in the sense of \cite{segal1987,segal1990} (see Appendix VIII of \cite{ahn2014}). Like the kinked and smooth specifications, the generalized kinked specification is FOSD-rationalizable under multiple priors but not necessarily under a single prior.}

To conclude, the kinked and smooth specifications in equations (\ref{eq:kinked}) and (\ref{eq:smooth}) both encode ambiguity aversion by tilting weight toward the uncertain state where demand is lower---the smaller of $x_1$ and $x_3$. What makes this possible is that both treat the two securities paying off in the uncertain states symmetrically, so that the weight attached to each depends only on the ranking of $x_1$ and $x_3$ and never on the label of the state. In the kinked specification this symmetry is enforced by the $\min$ and $\max$ operators, and in the smooth specification it follows from the symmetry of $\mu$ about $\epsilon = 0$. Because the identity of this state---and hence the implied weighting---shifts with the prices and the resulting allocation, no single prior can reproduce this utility pattern. This is what the Ellsberg-type example in the Introduction reveals, where the choices imply that $\pi_1 < \tfrac{1}{3}$ in one budget and $\pi_3 < \tfrac{1}{3}$ in another, contradicting $\pi_1 + \pi_3 = \tfrac{2}{3}$.

\section{Revealed Preference Analysis} \label{sec:analysis}

Our empirical analysis is nonparametric and rests entirely on revealed preference: for each subject we ask which of several nested rationalizability concepts, from basic rationalizability to EUT-rationalizability, is compatible with the observed choices, and by how much. Section \ref{subsec:concepts} defines these rationalizability concepts and Section \ref{subsec:scores} turns them into cost-efficiency scores that gauge approximate compatibility. The framework extends \cite{dembo2026} from risk to uncertainty so we keep the expositional treatment relatively brief, referring to the Appendix and to \cite{dembo2026} for details on the revealed preference tests and the formal results that justify their use. That said, there are two rationalizability concepts formulated and tested in this paper which are not found in \cite{dembo2026}: FOSD-rationalizability under multiple priors, which accommodates ambiguity aversion (and seeking), and rationalizability of the kinked specification.

\subsection{Rationalizability Concepts} \label{subsec:concepts}

A subject's data are the collection of price-allocation pairs $\cal{D} = \{(\mbp^i, \mbx^i)\}_{i = 1}^{50}$ described in Section \ref{sec:data}. We refer to a utility function $U : \mbbR^3_+ \to \mbbR$ as \emph{well-behaved} if it is continuous and increasing.\footnote{By increasing, we mean that $U(\mbx') > U(\mbx)$ whenever $\mbx' > \mbx$; note that $\mbx' \geq \mbx$ if $x_s' \geq x_s$ for all $s$, and $\mbx' > \mbx$ if $\mbx' \geq \mbx$ and $\mbx' \neq \mbx$.} We consider four families of well-behaved utility functions, each nested within the last, and then the kinked and smooth specifications of Section \ref{sec:models}.

\subsubsection{Basic Rationalizability} \label{subsubsec:basic-rat}

A dataset $\cal{D}$ is rationalizable if it is compatible with some well-behaved utility function $U$, in the sense that $U(\mbx^i) \geq U(\mbx)$ for all $\mbx \in \cal{B}^i = \{\mbx \in \mbbR^3_+ : \mbp^i \cdot \mbx \leq 1\}$. By \citeauthor{afriat1967}'s (\citeyear{afriat1967}) theorem, $\cal{D}$ is rationalizable if and only if it satisfies the generalized axiom of revealed preference (GARP), which requires that $\cal{D}$ is free of revealed preference cycles.\footnote{Given $\cal{D} = \{(\mbp^i, \mbx^i)\}_{i = 1}^{50}$, let $\cal{X} = \{\mbx^i\}_{i = 1}^{50}$. For any allocations $\mbx^i$ and $\mbx^j$ in $\cal{X}$, we say that $\mbx^i$ is \emph{directly revealed preferred (directly revealed strictly preferred)} to $\mbx^j$ if $\mbp^i \cdot \mbx^j \leq (<) \, \, 1$. A \emph{revealed preference cycle} occurs if there are allocations $\mbx^{i_1}$, $\mbx^{i_2}, \ldots, \mbx^{i_n}$ in $\cal{X}$, with $\mbx^{i_n} = \mbx^{i_1}$, such that $\mbx^{i_k}$ is directly revealed preferred to $\mbx^{i_{k + 1}}$, for $k = 1, 2, \ldots, n - 1$, and at least one of the direct revealed preference relations is strict.}

A weaker version of GARP is the weak axiom of revealed preference (WARP), which requires that $\cal{D}$ cannot have pairwise revealed preference cycles.\footnote{To be precise, there are no allocations $\mbx^i$ and $\mbx^j$ in $\cal{X}$ such that each allocation is directly revealed preferred to the other and one of the direct revealed preference relations is strict.} It is straightforward to check that the maximization of a \emph{complete} preference subject to a budget constraint implies that $\cal{D}$ obeys WARP (see Appendix I of \cite{dembo2026}). Note, however, that such a preference need not be \emph{transitive} and so may not have a utility representation. Thus a violation of WARP implies a violation of completeness; on the other hand, if $\cal{D}$ violates GARP but respects WARP, then $\cal{D}$ is rationalizable by a complete preference, but that preference cannot be complete and transitive.

It is well known that when there are just two goods, WARP and GARP are observationally indistinguishable in the sense that a dataset satisfies one property if and only if it satisfies the other (see \cite{rose1958}). The three-dimensional budgets in our experiment make GARP and WARP observationally distinct and therefore we can separate failures of completeness from failures of transitivity.

\subsubsection{FOSD-Rationalizability} \label{subsubsec:fosd-rat}

Between the basic rationalizability and EUT-rationalizability concepts lies rationalizability by a family of utility functions that are monotone with respect to first-order stochastic dominance (FOSD): a well-behaved utility function $U$ is FOSD-increasing if $U(\mbx'') \geq U(\mbx')$ whenever the payoff distribution induced by $\mbx''$ first-order stochastically dominates that induced by $\mbx'$, and strictly when the dominance is strict. Violations of FOSD are naturally interpreted as errors rather than as expressions of risk or ambiguity attitudes, which is why this monotonicity condition is widely imposed. Recall that state 2 has the objectively known probability $\tfrac{1}{3}$, while the uncertain states 1 and 3 have unknown probabilities that sum to $\tfrac{2}{3}$; because the latter are unknown, FOSD-rationalizability takes two forms:
\begin{itemize}
\item \textbf{Multiple Priors} \; A dataset $\cal{D}$ is \emph{FOSD-rationalizable under multiple priors} if it is compatible with a well-behaved utility function $U$ that is ($i$) symmetric across the two uncertain states, and ($ii$) FOSD-increasing given the objectively known probabilistic structure (state 2 occurs with probability $\tfrac{1}{3}$, and states 1 and 3 share the remaining $\tfrac{2}{3}$). FOSD-rationalizability under multiple priors accommodates ambiguity aversion and seeking and contains the kinked and smooth specifications of Section \ref{sec:models}.
\item \textbf{Uniform Prior} \; A dataset $\cal{D}$ is \emph{FOSD-rationalizable under the uniform prior} if it is compatible with a well-behaved utility function $U$ that is FOSD-increasing with respect to the uniform prior, or equivalently, increasing and symmetric across all three states. Note that in the experiment conducted by \cite{dembo2026}, the distribution over the three states is commonly known to be uniform, and what \cite{dembo2026} simply refers to as ``FOSD-rationalizability'' corresponds exactly to what we refer to as ``FOSD-rationalizability under the uniform prior.''
\end{itemize}

FOSD-rationalizability under the uniform prior is the more demanding of the two rationalizability concepts. The gap between the two concepts is precisely what the canonical non-EUT models of ambiguity exploit: they are FOSD-rationalizable under multiple priors but not necessarily under the uniform prior. We test both notions using the results of \cite{nishimura2017}, which strengthen GARP to rule out a larger class of revealed preference cycles; see the Appendix for details.

We have defined FOSD-rationalizability under the uniform prior $\mbpi = (\tfrac{1}{3}, \tfrac{1}{3}, \tfrac{1}{3})$, and the EUT-rationalizability concept below is also stated under the uniform prior. Both rationalizability concepts extend to any single prior $\mbpi = (\pi_1, \tfrac{1}{3}, \pi_3)$ such that $\pi_1 + \pi_3 = \tfrac{2}{3}$, and we analyze and present results for the full range of single priors in Section \ref{subsec:uniform}.

\subsubsection{EUT-Rationalizability} \label{subsubsec:eut-rat}

The most stringent rationalizability concept we consider is rationalizability by subjective EUT. Under the uniform prior, subjective EUT requires a continuous and increasing Bernoulli index $u : \mbbR_+ \to \mbbR$ over monetary consumption with $U(\mbx) = \tfrac{1}{3} \sum_{s = 1}^3 u(x_s)$, and a dataset $\cal{D}$ is EUT-rationalizable if it is compatible with a well-behaved utility function $U$ of this form. Every such $U$ is FOSD-increasing and symmetric, so EUT-rationalizability is more demanding than FOSD-rationalizability under the uniform prior, which is in turn more demanding than FOSD-rationalizability under multiple priors. We test EUT-rationalizability using the generalized restriction of infinite domains (GRID) method of \cite{polisson2020}; this method is also used in \cite{dembo2026}.\footnote{Other revealed preference approaches for testing EUT-rationalizability include \cite{diewert2012}, \cite{bayer2013}, \cite{echenique2015}, \cite{chambers2016b}, and \cite{echenique2021}.} Using the GRID method, EUT-rationalizability can be tested by checking whether there is a solution to a finite system of linear inequalities, where the unknowns are the values of $u$ (the Bernoulli index) at a finite set of consumption levels.

\subsubsection{Kinked-Rationalizability and Smooth-Rationalizability} \label{subsubsec:kinked-rat}

Finally, we consider rationalizability by the kinked and smooth specifications of Section \ref{sec:models}. A dataset $\cal{D}$ is rationalizable by the kinked specification if it is compatible with a well-behaved utility function of the form in equation (\ref{eq:kinked}) for some $\alpha \in (0,1)$ and some Bernoulli index $u$, and by the smooth specification if it is compatible with one of the form in equation (\ref{eq:smooth}) for some $u$, $\varphi$, and $\mu$. Both specifications nest subjective EUT (at $\alpha = \tfrac{1}{2}$ and for linear $\varphi$, respectively) and both are FOSD-increasing under multiple priors (see Section \ref{sec:models}), so each concept lies between EUT-rationalizability and FOSD-rationalizability under multiple priors. Neither is nested with FOSD-rationalizability under the uniform prior, however: away from the EUT case, neither specification is symmetric across all three states, and a utility function that is symmetric need not take either form.

We test the kinked but not the smooth specification, for the reason given in Section \ref{sec:models}. For fixed $\alpha$, equation (\ref{eq:kinked}) has known weights on $x_{\min}$, $x_2$, and $x_{\max}$ and an unknown $u$. The GRID method of \cite{polisson2020} remains applicable and the test again involves checking whether there is a solution to a finite system of linear inequalities, where the unknowns are the values of $u$ at a finite set of consumption levels. Repeating the test across $\alpha$ leaves the ambiguity parameter unrestricted. In equation (\ref{eq:smooth}), by contrast, the restrictions on choice depend jointly on $u$, $\varphi$, and $\mu$, and because $\varphi$ acts on expected utility levels they remain nonlinear in $u$ even for fixed $\varphi$ and $\mu$. A feasible test would have to restrict $\varphi$ and $\mu$ parametrically and would not be comparable to the other tests. Since the smooth specification is FOSD-increasing under multiple priors, however, its performance is bounded by that of the multiple-prior class (see Section \ref{subsec:scores}).

\subsection{Rationalizability Scores} \label{subsec:scores}

A subject's data need not be exactly rationalizable by any of the rationalizability concepts outlined in Section \ref{subsec:concepts}, so we measure how close they come using the critical cost-efficiency index (CCEI) of \cite{afriat1972,afriat1973}. For $e \in (0,1]$, a dataset $\cal{D}$ is rationalizable at cost-efficiency $e$ if it is compatible with some well-behaved utility function $U$ in the relevant family once each budget is shrunk to $\cal{B}^i(e) = \{\mbx \in \mbbR^3_+ : \mbp^i \cdot \mbx \leq e\}$, and the CCEI is the largest such $e$. For example, $\estar$ is the supremum of those values of $e$ such that there is a well-behaved utility function $U$ with $U(\mbx^i) \geq U(\mbx)$ for all $\mbx \in \cal{B}^i(e)$. The CCEI has the usual money-metric reading: a subject leaves up to the fraction one minus the CCEI of each budget on the table through inconsistent choices. We refer the interested reader to the Appendix for a fuller explanation of how the CCEI for different families of utility functions could be calculated.

Computing the CCEI for each of the four nested rationalizability concepts of Sections \ref{subsubsec:basic-rat} to \ref{subsubsec:eut-rat} gives four scores per subject: $\estar$ for basic rationalizability, $\tildeedoublestar$ for FOSD-rationalizability under multiple priors, $\edoublestar$ for FOSD-rationalizability under the uniform prior, and $\etriplestar$ for EUT-rationalizability. Since each family is contained within the previous one, the scores are nested:
\[
1 \geq \estar \geq \tildeedoublestar \geq \edoublestar \geq \etriplestar > 0,
\]
so that each consecutive gap measures the additional cost of imposing the corresponding requirement, placing every subject's departure from EUT-rationalizability on a single scale.

The kinked specification yields a fifth score. For each $\alpha \in (0,1)$, we can compute the CCEI for the family of utility functions given by equation (\ref{eq:kinked}). Denoting this value by $e^{\kappa}(\alpha)$, the kinked-specification score is the best fit over the ambiguity parameter: $e^{\kappa} = \max_{\alpha} e^{\kappa}(\alpha)$. Because the specification reduces to subjective EUT at $\alpha = \tfrac{1}{2}$, we have $e^{\kappa}(\tfrac{1}{2}) = \etriplestar$, and so $e^{\kappa} \geq \etriplestar$; because every kinked utility function is FOSD-increasing under multiple priors, any budget shrinkage that rationalizes the data by the kinked specification also rationalizes the data under multiple priors, so $e^{\kappa} \leq \tildeedoublestar$. No such ordering holds relative to $\edoublestar$: a kinked utility function with $\alpha \neq \tfrac{1}{2}$ is not symmetric across all three states, and a utility function that is need not be kinked, so $e^{\kappa}$ may lie on either side of $\edoublestar$. Which is larger---whether a subject's choices are better explained by the kinked specification or by probabilistic sophistication---is an empirical question, and the improvement over subjective EUT is $\max \, \{e^{\kappa}, \edoublestar\} - \etriplestar$. Since $\tildeedoublestar$ bounds the performance of the entire multiple-prior class, kinked and smooth alike, $\tildeedoublestar - e^{\kappa}$ bounds what any other multiple-prior model, including the smooth specification, could add over and above the kinked specification.

Finally, we note that the CCEI is not the only way to gauge how far a subject's choices fall short of exact rationalizability. Other efficiency-based indices exist, and \cite{dembo2026} also develops an alternative, distance-based measure, which instead asks how far the observed choices would have to be moved on the budget constraint to be rationalized. We use the CCEI for the reasons \cite{dembo2026} gives: it has a natural economic interpretation, it is the standard measure in the revealed preference literature, and it can be computed for each of our nested concepts, whereas no exact procedure is known for calculating the distance-based measure for EUT-rationalizability. The CCEI and distance-based measure deliver broadly the same conclusions in any case. We refer the reader to \cite{dembo2026} for the comparison and details.

\section{Illustrative Subjects} \label{sec:subjects}

The dataset from the experiment of \cite{ahn2014} contains observations on 154 individual subjects. For each subject, we have 50 observations $\cal{D} = \{(\mbp^i, \mbx^i)\}_{i = 1}^{50}$, where $\mbp^i = (p_1^i, p_2^i, p_3^i)$ denotes the $i$-th observation of the price vector and $\mbx^i = (x_1^i, x_2^i, x_3^i)$ denotes the corresponding demand allocation. Recall that state 2 is risky and has a $\tfrac{1}{3}$ chance of occurring, and states 1 and 3 are uncertain and occur according to unknown probabilities summing to $\tfrac{2}{3}$.

\cite{ahn2014} provides an overview of the important aggregate features of the data, but these can tell us little about the choice behavior of individual subjects. To get a better sense of the wide range of observed behavior, we next provide an overview of some key features of the individual-level data. We focus on subjects whose behavior corresponds to one of several prototypical attitudes toward risk and/or uncertainty to illustrate the regularity within subjects and heterogeneity across subjects that is characteristic of the data. These selected cases also bring out the heterogeneity in subjects' consistency with the different rationalizability concepts. We also report and discuss these subjects' rationalizability scores, which helps to build intuition for how the scores work and how they pinpoint the sources of departure from EUT-rationalizability. Note that these are special cases where the regularities in the data are particularly clear; scatterplots for the full set of subjects are available in Appendix III of \cite{ahn2014}.

Figure \ref{fig:figure2} depicts the allocations chosen by six individual subjects, expressed either as token shares or budget shares across the three securities, as points in the unit simplex. The vertices of the unit simplex correspond to allocations consisting entirely of one of the three securities. Each interior point represents an allocation as a convex combination of the vertices. For any allocation $\mbx^i = (x_1^i, x_2^i, x_3^i)$, we define the \emph{token share} of the security that pays off in state $s$ to be the number of tokens payable in state $s$ as a fraction of the sum of tokens payable in all three states $x_s^i/(x_1^i + x_2^i + x_3^i)$. We also define the \emph{budget share} (or expenditure share) of the security that pays off in state $s$ to be the expenditure on tokens invested in this security as a fraction of total expenditure; since prices are normalized so that total expenditure equals unity, the budget share is simply $p_s^i x_s^i$.

Before proceeding to the scatterplots, we tabulate the CCEI-type scores for each of the six illustrative subjects whose choices are depicted in Figure \ref{fig:figure2}: $\estar$ for (basic) rationalizability, $\tildeedoublestar \leq \estar$ for FOSD-rationalizability under multiple priors, and $\etriplestar \leq \tildeedoublestar$ for EUT-rationalizability. We do not report the uniform-prior FOSD-rationalizability scores $\edoublestar \leq \tildeedoublestar$ for these six subjects, as they are all roughly equal (exactly equal for five subjects) to their EUT-rationalizability scores $\etriplestar$. We also report the score $e^{\kappa}$ for the kinked specification of Section \ref{sec:models}, interpreted through the lens of the $\alpha$-MEU model, which lies between the EUT and multiple-prior scores, $\etriplestar \leq e^{\kappa} \leq \tildeedoublestar$. It is set apart in the table because, unlike the other three, it measures the performance of one particular specification rather than bounding a whole class of models; its construction is given in Section \ref{subsec:scores}. Columns (a)--(f) correspond to the panels of Figure \ref{fig:figure2}.
\begin{center}
\begin{tabular}{r|c|c|c|c|c|c}
& (a) & (b) & (c) & (d) & (e) & (f) \\
\hline\hline
$\estar$ & 0.994 & 1.000 & 1.000 & 0.943 & 0.981 & 0.934 \\
\hline
$\tildeedoublestar$ & 0.966 & 0.994 & 0.999 & 0.928 & 0.952 & 0.848 \\
\hline
$\etriplestar$ & 0.965 & 0.994 & 0.991 & 0.660 & 0.801 & 0.750 \\
\hline\hline
$e^{\kappa}$ & 0.966 & 0.994 & 0.993 & 0.852 & 0.871 & 0.848
\end{tabular}
\end{center}
Note that the choices of all six illustrative subjects are nearly rationalizable ($\estar \approx 1$), but only the first three subjects, whose choices are depicted in panels (a)--(c), are also nearly EUT-rationalizable ($\etriplestar \approx 1$). The subjects whose choices are depicted in panels (d)--(f) are rationalizable but display meaningful departures from EUT-rationalizability.

\begin{figure}[!t]
\begin{center}
\subfloat[ID 703]{\includegraphics[scale=.45]{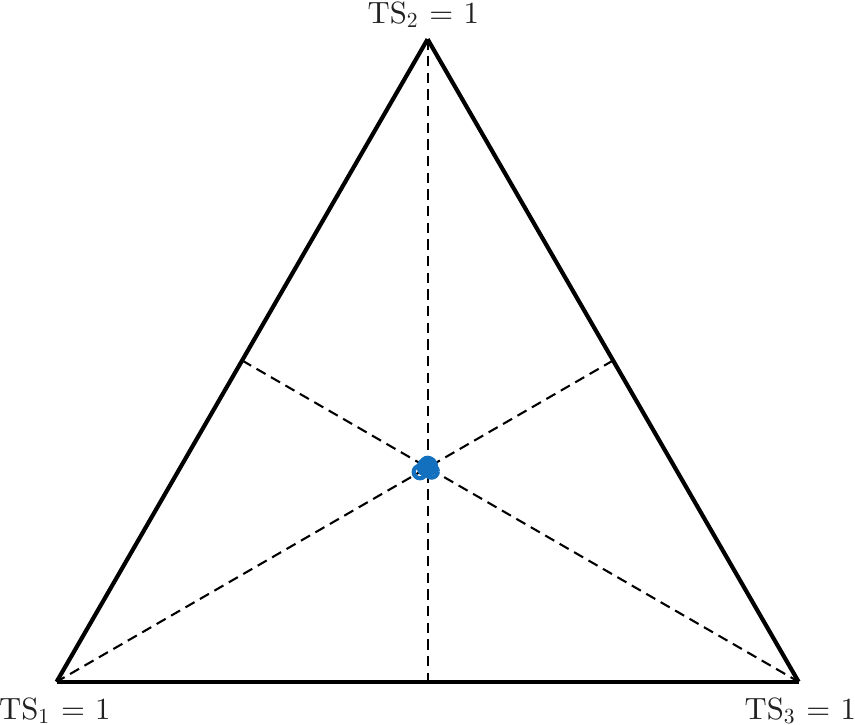}} \hspace{.25in} \subfloat[ID 629]{\includegraphics[scale=.45]{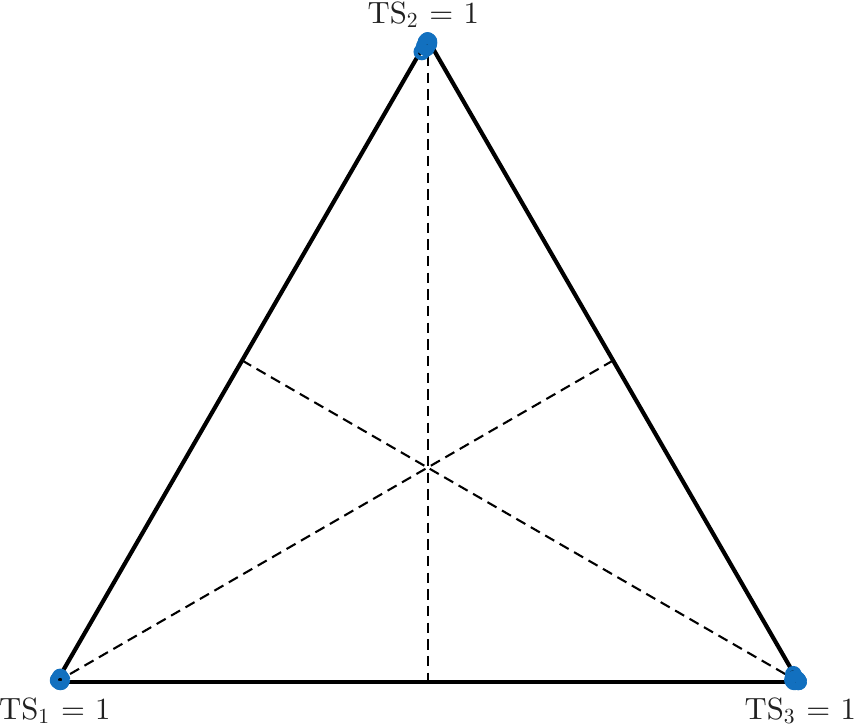}} \\
\subfloat[ID 921]{\includegraphics[scale=.45]{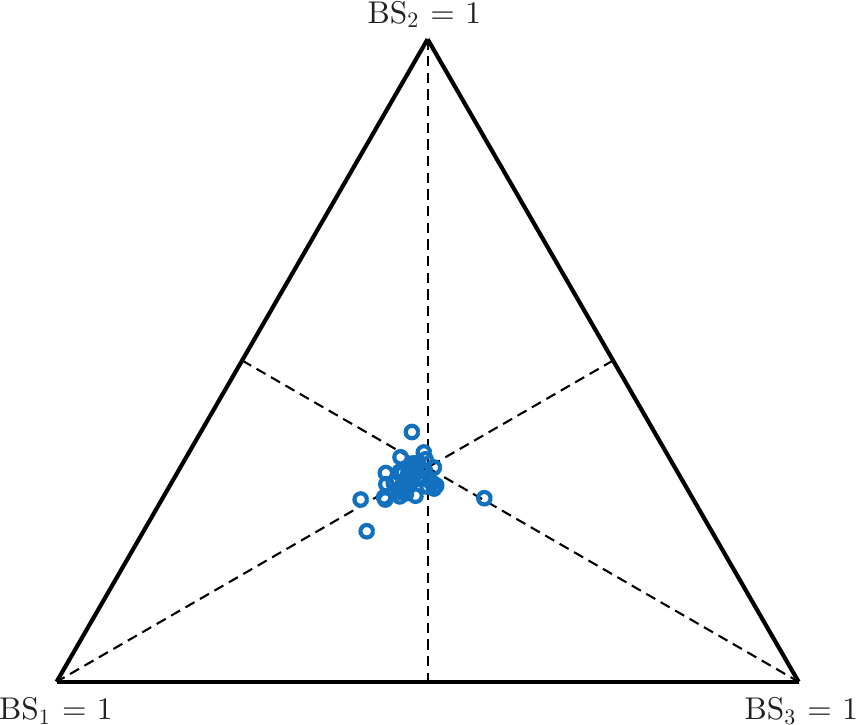}} \hspace{.25in} \subfloat[ID 340]{\includegraphics[scale=.45]{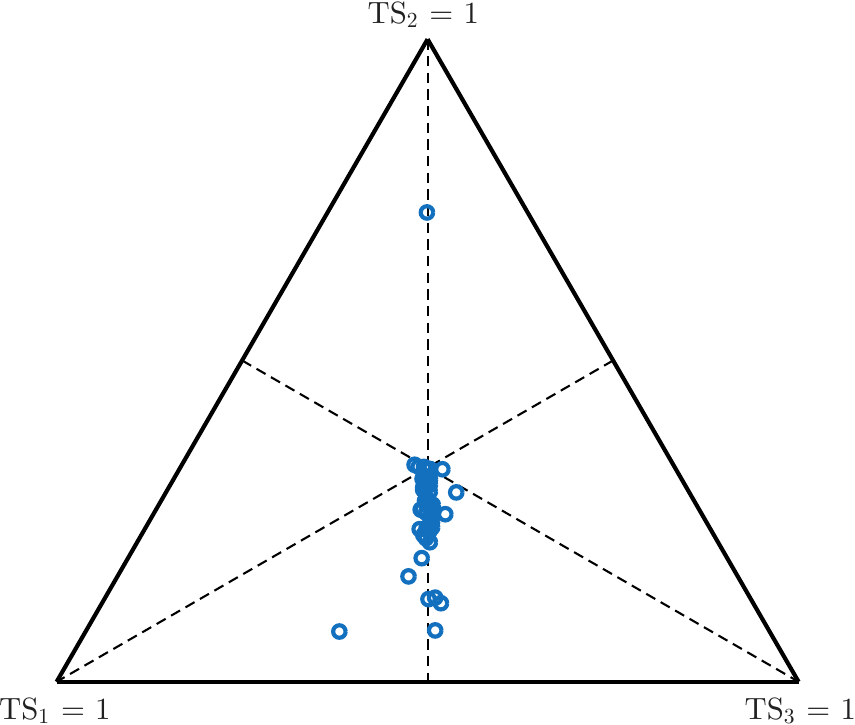}} \\
\subfloat[ID 407]{\includegraphics[scale=.45]{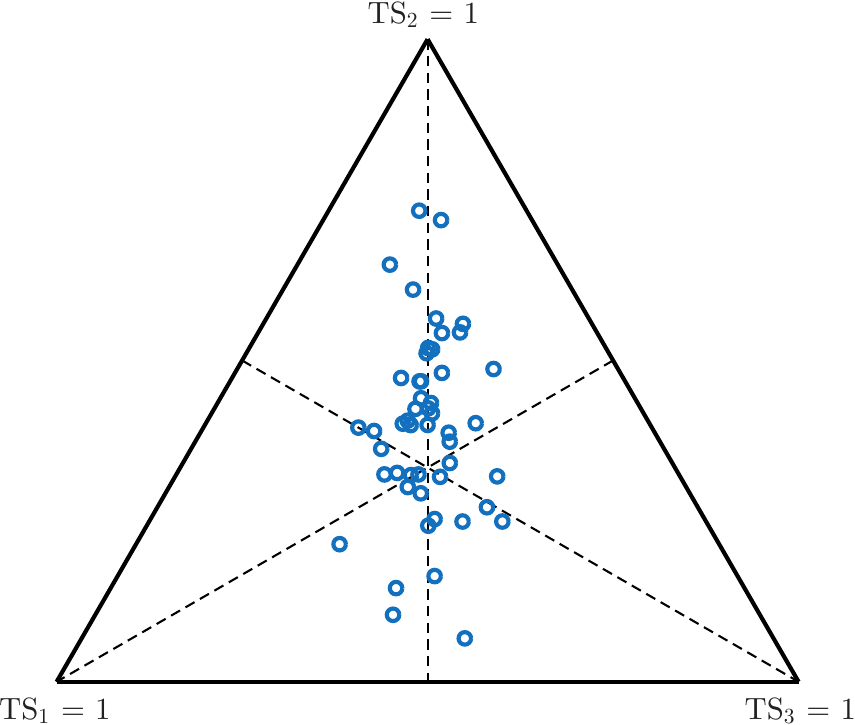}} \hspace{.25in} \subfloat[ID 612]{\includegraphics[scale=.45]{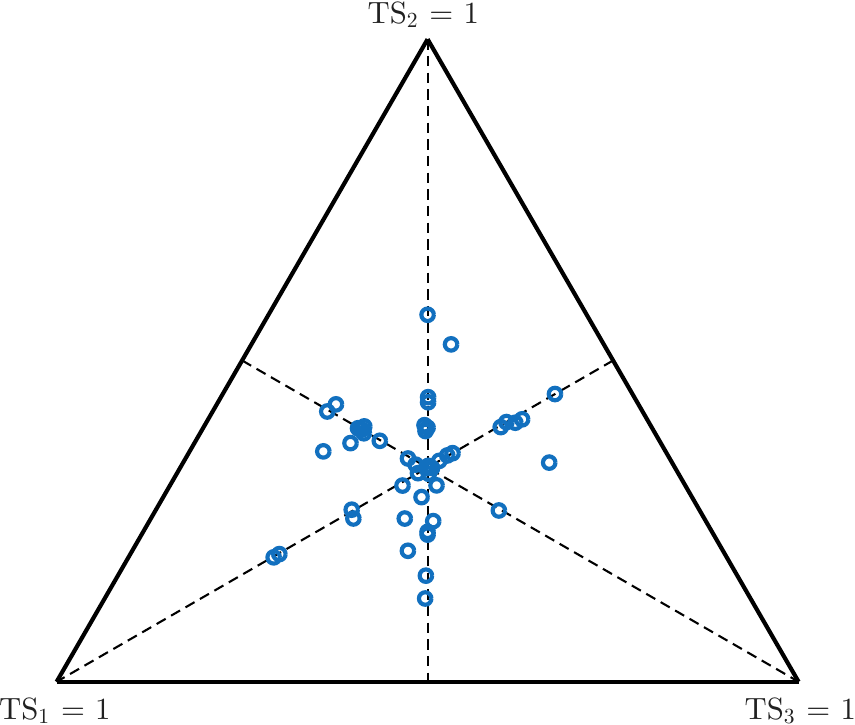}}
\end{center}
\captionsetup{width=.95\linewidth}
\vspace{-.1in}
\caption{Subject Behavior}
\vspace{-.1in}
\caption*{\footnotesize{Each panel shows all 50 choices for a single subject in terms of token shares, except for panel (c) which is in terms of budget shares. Each vertex of the unit simplex corresponds to a full allocation to one of the three securities. (a) ID 703 is consistent with infinite risk aversion; (b) ID 629 is consistent with risk neutrality; (c) ID 921 is consistent with the maximization of logarithmic von Neumann-Morgenstern subjective expected utility; (d) ID 340 and (e) ID 407 are consistent with ambiguity aversion; and (f) ID 612 is consistent with pessimism.}}
\label{fig:figure2}
\end{figure}

In all panels of Figure \ref{fig:figure2}, choices are depicted in terms of token shares, except in panel (c), where they are depicted in terms of budget shares. Figure \ref{fig:figure2}a depicts the choices of a subject (ID 703) who always chooses nearly equal amounts $x_1^i = x_2^i = x_3^i$, suggesting infinite risk aversion. Figure \ref{fig:figure2}b shows a very different case, the choices of a subject (ID 629) who, with a few exceptions, invests all his tokens in the cheapest security, behavior consistent with risk neutrality. Figure \ref{fig:figure2}c depicts the choices of a subject (ID 921) who equalizes expenditures $p_1^i x_1^i = p_2^i x_2^i = p_3^i x_3^i$, rather than tokens, across the three securities. This behavior is consistent with maximizing a logarithmic expected utility function.

Figure \ref{fig:figure2}d shows the choices of a subject (ID 340) who, with very few exceptions, allocates nearly equal amounts to the securities that pay off in the uncertain states $x_1^i = x_3^i \neq x_2^i$. Figure \ref{fig:figure2}e shows the choices of a subject (ID 407) who does not demand nearly equal amounts of the securities that pay off in the uncertain states $x_1^i \neq x_3^i$, but these demands are much closer to each other than to the demand for the security that pays off in the risky state $x_2^i$. The behavior of these subjects suggests ambiguity aversion, in the sense that they are trying to reduce the sensitivity of their payoffs to states with unknown probabilities.

The main evidence for ambiguity aversion in the data is the tendency for subjects to equate, or at least reduce the difference in, their demands for the securities that pay off in the uncertain states. However, there is a similar, though weaker, tendency to equate the demands for pairs of securities where one security pays off in an uncertain state and the other in the risky state, which indicates pessimism (overweighting the probability of low payoffs). Figure \ref{fig:figure2}f shows the choices of one such subject (ID 612), who also equalizes the demands for the two pairs of securities where exactly one security pays off in an uncertain state ($x_1^i = x_2^i$ and $x_2^i = x_3^i$).

The subjects depicted in Figure \ref{fig:figure2}d (ID 340) and Figure \ref{fig:figure2}e (ID 407) are almost as FOSD-rationalizable under multiple priors as they are rationalizable ($\tildeedoublestar \approx \estar$), but have substantially lower EUT-rationalizability scores ($\etriplestar < \tildeedoublestar$), consistent with strong ambiguity aversion reflected in the near-equalization of demands for the uncertain securities. For the subject depicted in Figure \ref{fig:figure2}f (ID 612), by contrast, departures from EUT-rationalizability are also driven by FOSD violations even under multiple priors ($\etriplestar < \tildeedoublestar < \estar$).

The kinked-specification scores split the gap between $\tildeedoublestar$ and $\etriplestar$ into the part that one particular multiple-prior specification captures and the part it leaves unexplained. For the subjects depicted in panels (a)--(c) there is nothing to split: $\etriplestar$ is already close to $\estar$, so no non-EUT model has room to improve, and $e^{\kappa}$ is virtually identical to $\etriplestar$. The subjects depicted in panels (d)--(f) are more instructive. For the subject depicted in Figure \ref{fig:figure2}d (ID 340), whose near-equalization of the uncertain securities is the canonical hedging pattern, the kinked specification delivers the bulk of the available improvement, $e^{\kappa} - \etriplestar = 0.193$ against a total scope of $\tildeedoublestar - \etriplestar = 0.269$, and the residual $\tildeedoublestar - e^{\kappa} = 0.076$ bounds what any other multiple-prior model could add for this subject.\footnote{Differences between scores are computed before rounding, so they need not coincide with the differences between the rounded entries in the table above.} For the subject depicted in Figure \ref{fig:figure2}e (ID 407), who hedges only partially, the improvement over subjective EUT and the residual are of comparable size (0.070 and 0.080), so the kinked specification captures roughly half of what the multiple-prior class can offer. For the subject depicted in Figure \ref{fig:figure2}f (ID 612), $e^{\kappa}$ and  $\tildeedoublestar$ agree to three decimal places (0.848), so the kinked specification exhausts the multiple-prior scope entirely: the whole improvement over subjective EUT, $e^{\kappa} - \etriplestar = 0.099$, is delivered by the kinked specification alone. That this subject nonetheless remains well short of $\estar$ is a reminder that the two gaps measure different things: the departures that no multiple-prior model can absorb are the FOSD violations recorded in $\estar - \tildeedoublestar$, not in the gap over which the kinked specification competes. This heterogeneity underscores that different subjects deviate differently from the axioms underlying models of choice under uncertainty, which is precisely why our comprehensive individual-level nonparametric tests are important: pooling data or focusing on a single axiom would obscure these distinct patterns of behavior and mischaracterize the sources of departure.

\section{Experimental Results} \label{sec:results}

In this section, we present our empirical results. Throughout, we report the results from the uncertainty experiment of \cite{ahn2014} alongside the corresponding results from the risk experiment of \cite{dembo2026}, so that every finding under uncertainty can be read directly against its counterpart under risk. The section proceeds from coarse to fine, then to robustness, and finally to a test of a specific model. Section \ref{subsec:aggregate} draws the central comparison between the two experiments: we first summarize the underlying WARP and GARP violations and the CCEI scores they generate, and then compare the full distributions of the scores for rationalizability ($\estar$), FOSD-rationalizability under the uniform prior ($\edoublestar$), and EUT-rationalizability ($\etriplestar$) across the two decision domains; we also discuss FOSD-rationalizability under multiple priors in the domain of uncertainty ($\tildeedoublestar$). Section \ref{subsec:uniform} asks whether our findings depend on the uniform prior we adopt under uncertainty, and shows that the uniform prior approximately maximizes single-prior rationalizability, so that no other single prior would materially change our conclusions. Section \ref{subsec:disaggregate} moves from the aggregate distributions to the individual level, locating the source of each subject's departure from EUT-rationalizability and testing, subject by subject, the magnitudes of the gaps between scores. Section \ref{subsec:power} evaluates the power of our tests by comparing the observed scores to those generated from simulated data. Finally, Section \ref{subsec:kinked} complements this model-free analysis with a test of one particular model: it computes rationalizability scores for the kinked specification, interpreted through the lens of $\alpha$-MEU, and asks how much of the scope for multiple-prior models identified in the preceding subsections that specification captures.

\subsection{Rationalizability under Uncertainty versus Risk} \label{subsec:aggregate}

First, we compare the rationalizability of choices under uncertainty with the rationalizability of choices under risk. We begin with the most basic revealed preference diagnostics---the WARP and GARP violations that underlie the CCEI rationalizability scores---and then compare the distributions of rationalizability scores across the two experiments. Throughout, we take the uniform prior $\mbpi = (\tfrac{1}{3}, \tfrac{1}{3}, \tfrac{1}{3})$ as the natural single-prior benchmark for the uncertainty experiment, given its symmetric design; Section \ref{subsec:uniform} confirms that this choice is innocuous. Under the uniform prior, the scores are nested: $1 \geq \estar \geq \tildeedoublestar \geq \edoublestar \geq \etriplestar > 0$.

Before turning to the rationalizability scores, we describe the underlying revealed preference violations exhibited by individual subjects. Table \ref{tab:table1} reports percentile values of the number of WARP violations, GARP violations, and GARP violations that do not contain a WARP violation, alongside the CCEI scores required to remove all violations of WARP and GARP. The upper panel of Table \ref{tab:table1} reports the results under uncertainty and the lower panel under risk. The number of GARP violations is the number of distinct revealed preference cycles; a WARP violation is a special case of a GARP violation involving a pairwise cycle.

\begin{table}[!t]
\begin{center}
\begin{tabular}{|r|r|c|c|c||c|c|}
\hline
\multicolumn{7}{|c|}{Uncertainty} \\
\hline
\hline
\multicolumn{2}{|c|}{} & \multicolumn{3}{c||}{Violations} &  \multicolumn{2}{c|}{$1-\text{CCEI}$} \\
\cline{3-7}
\multicolumn{2}{|c|}{} & \multirow{2}{*}{WARP} & \multirow{2}{*}{GARP} & GARP &  \multirow{2}{*}{WARP} &  \multirow{2}{*}{GARP} \\
\multicolumn{2}{|c|}{} & & & not WARP & & \\
\cline{3-7}
\hline
\multirow{9}{*}{\rotatebox[origin=c]{90}{Percentile Values}} & 1 & 0 & 0 & 0 & 0.000 & 0.000 \\
\cline{2-7}
& 5 & 0 & 0 & 0 & 0.000 & 0.000 \\
\cline{2-7}
& 10 & 0 & 0 & 0 & 0.000 & 0.000 \\
\cline{2-7}
& 25 & 2 & 2 & 0 & 0.009 & 0.009 \\
\cline{2-7}
& 50 & 4 & 10 & 0 & 0.036 & 0.036 \\
\cline{2-7}
& 75 & 9 & 262 & 0 & 0.076 & 0.076 \\
\cline{2-7}
& 90 & 16 & 111,506 & 2 & 0.134 & 0.136 \\
\cline{2-7}
& 95 & 21 & 9,057,541 & 6 & 0.163 & 0.163 \\
\cline{2-7}
& 99 & 52 & $\geq 10^7$ & 235 & 0.238 & 0.238 \\
\hline
\multicolumn{7}{c}{ } \\
\hline
\multicolumn{7}{|c|}{Risk} \\
\hline
\hline
\multicolumn{2}{|c|}{} & \multicolumn{3}{c||}{Violations} &  \multicolumn{2}{c|}{$1-\text{CCEI}$} \\
\cline{3-7}
\multicolumn{2}{|c|}{} & \multirow{2}{*}{WARP} & \multirow{2}{*}{GARP} & GARP &  \multirow{2}{*}{WARP} &  \multirow{2}{*}{GARP} \\
\multicolumn{2}{|c|}{} & & & not WARP & & \\
\cline{3-7}
\hline
\multirow{9}{*}{\rotatebox[origin=c]{90}{Percentile Values}} & 1 & 0 & 0 & 0 & 0.000 & 0.000 \\
\cline{2-7}
& 5 & 0 & 0 & 0 & 0.000 & 0.000 \\
\cline{2-7}
& 10 & 0 & 0 & 0 & 0.000 & 0.000 \\
\cline{2-7}
& 25 & 1 & 1 & 0 & 0.004 & 0.004 \\
\cline{2-7}
& 50 & 4 & 7 & 0 & 0.031 & 0.032 \\
\cline{2-7}
& 75 & 8 & 318 & 0 & 0.084 & 0.087 \\
\cline{2-7}
& 90 & 14 & 181,655 & 1 & 0.172 & 0.172 \\
\cline{2-7}
& 95 & 25 & $\geq 10^7$ & 7 & 0.213 & 0.213 \\
\cline{2-7}
& 99 & 55 & $\geq 10^7$ & 24 & 0.332 & 0.332 \\
\hline
\end{tabular}
\end{center}
\captionsetup{width=.95\linewidth}
\vspace{-.1in}
\caption{Number of WARP/GARP Violations and CCEI Scores (Uncertainty vs. Risk)}
\vspace{-.1in}
\caption*{\footnotesize{Percentile values of the number of WARP violations, GARP violations, and GARP violations that do not contain a WARP violation, along with the CCEI scores required to eliminate all WARP and GARP violations. The lower panel (risk) is reproduced from \cite{dembo2026}.}}
\label{tab:table1}
\end{table}

As shown in the upper panel of Table \ref{tab:table1}, the median number of WARP violations is 4 and the median number of GARP violations is 10. For a few subjects the number of GARP violations is high, but the vast majority of these GARP violations also contain WARP violations: only 25 subjects (16.2 percent) have GARP violations that do not contain one or more WARP violations. Moreover, all 20 subjects (13.0 percent) with no WARP violations also have no GARP violations, so every subject who violates GARP also violates WARP.

For each subject, we can also calculate the CCEI measuring the amount by which each budget constraint must be reduced to remove all violations of WARP, which must be weakly greater than the usual CCEI (measuring the amount by which each budget constraint must be reduced to remove all violations of GARP). These two scores are identical for all but 4 subjects (2.6 percent), and as shown in the last two columns of the upper panel of Table \ref{tab:table1}, the distributions are nearly identical and concentrated near one.

Overall, departures from rationalizability are small---most subjects are close to being rationalizable---but where deviations occur, they arise from both incompleteness and nontransitivity: WARP violations (occurring whenever subjects depart from rationalizability) reflect incompleteness, while GARP violations that do not reduce to WARP violations (occurring for only a minority of subjects) reflect nontransitivity. Finally, for comparison, the lower panel of Table \ref{tab:table1} reproduces the analogous table of revealed preference violations and CCEI scores from the risk experiment of \cite{dembo2026}. The distributions of violations and scores across the two experiments are strikingly similar, which suggests that the introduction of uncertainty, over and above risk, does not materially degrade basic rationalizability---namely, the fundamental/conventional axioms of completeness and transitivity.

\begin{figure}[!t]
\begin{center}
\subfloat[$\estar$]{\includegraphics[scale=.37]{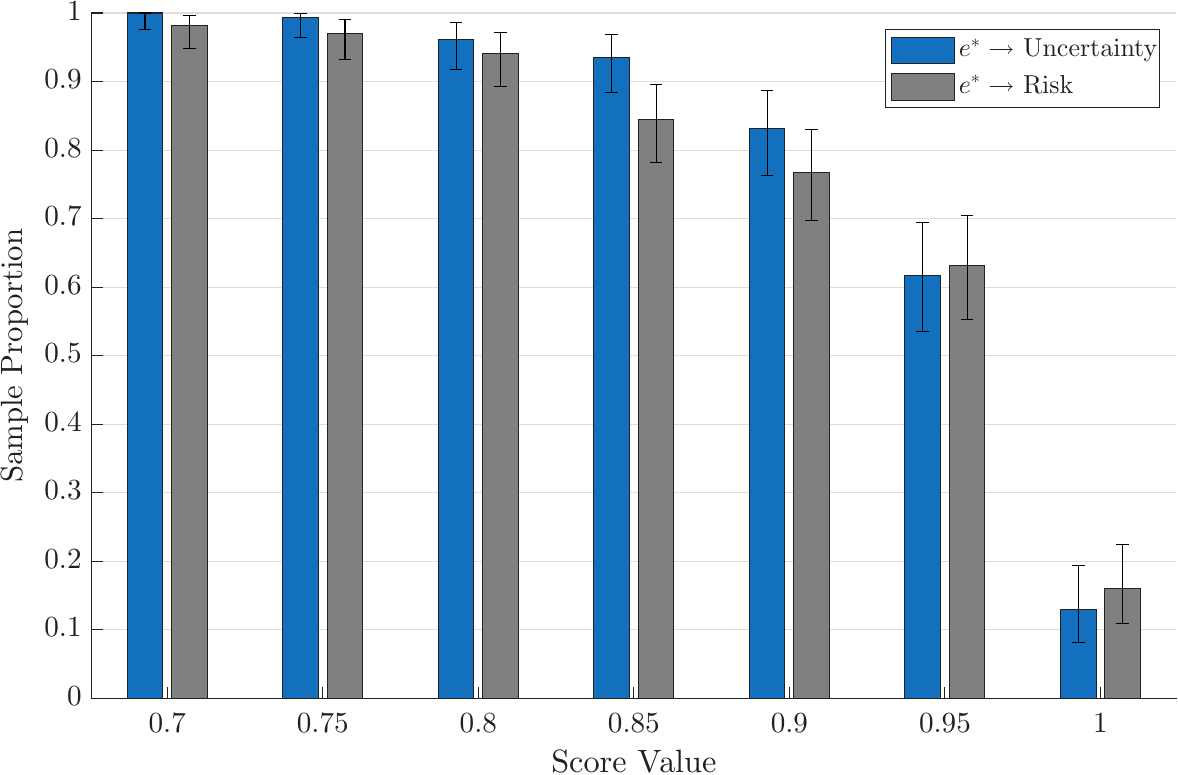}} \hspace{.5in} \subfloat[$\etriplestar$]{\includegraphics[scale=.37]{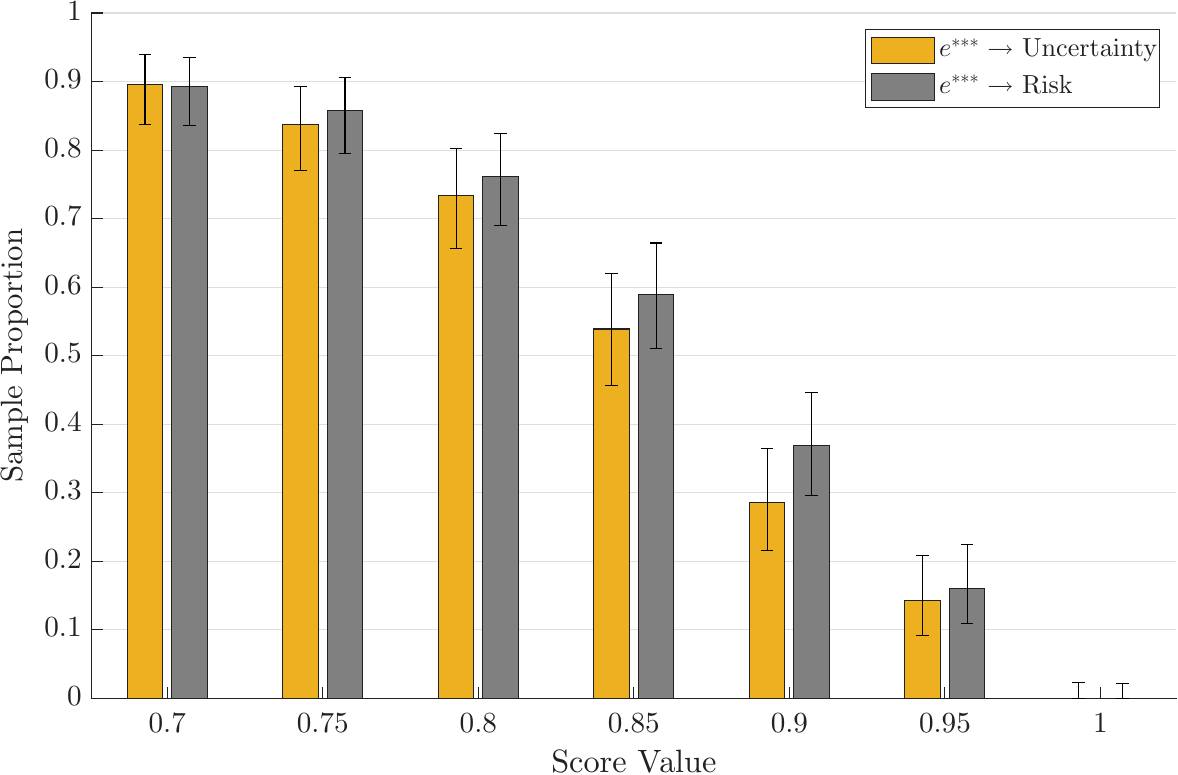}}
\end{center}
\captionsetup{width=.95\linewidth}
\vspace{-.1in}
\caption{Distributions of Rationalizability Scores (Uncertainty vs. Risk)}
\vspace{-.1in}
\caption*{\footnotesize{The horizontal axis shows the score values and the vertical axis presents the fraction of subjects with scores above each value for (a) rationalizability ($\estar$) and (b) EUT-rationalizability ($\etriplestar$). The braces indicate 95 percent confidence intervals on these proportions.}}
\label{fig:figure3}
\end{figure}

\begin{figure}[!t]
\begin{center}
\includegraphics[scale=.5]{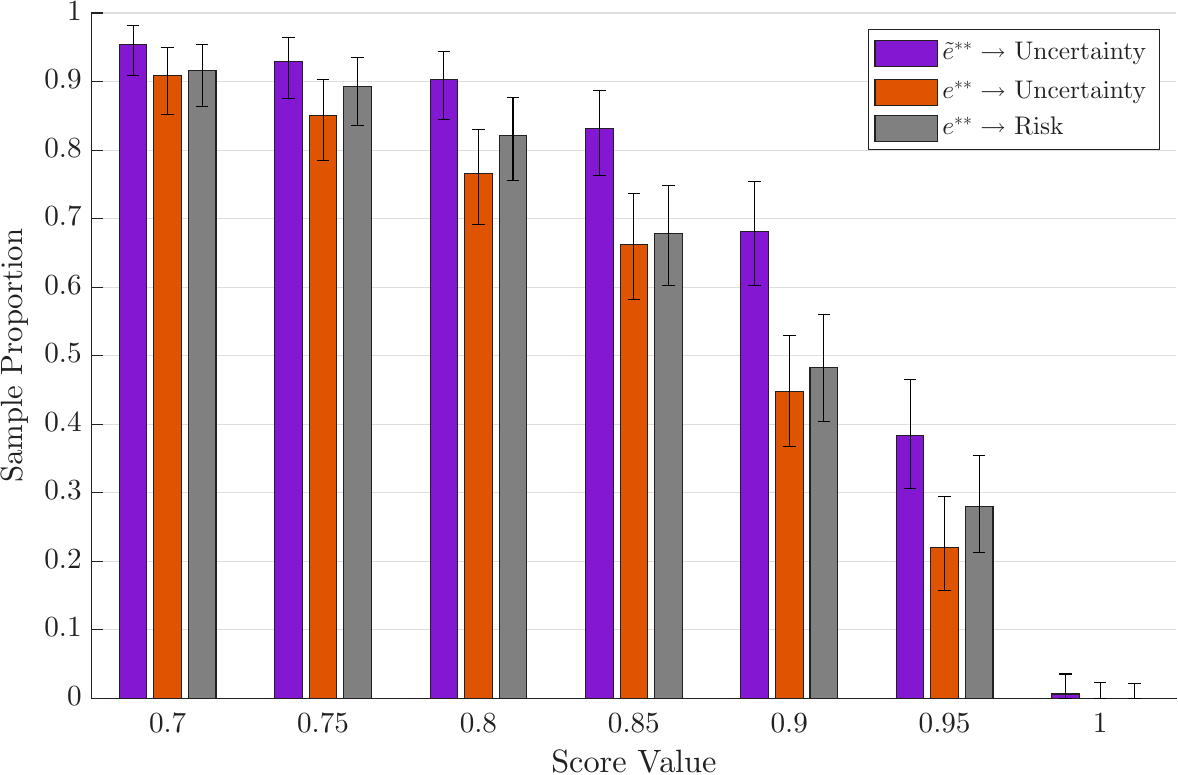}
\end{center}
\captionsetup{width=.95\linewidth}
\vspace{-.1in}
\caption{Distributions of FOSD-Rationalizability Scores (Uncertainty vs. Risk)}
\vspace{-.1in}
\caption*{\footnotesize{The horizontal axis shows the score values and the vertical axis presents the fraction of subjects with scores above each value for FOSD-rationalizability ($\tildeedoublestar$, $\edoublestar$). The braces indicate 95 percent confidence intervals on these proportions.}}
\label{fig:figure4}
\end{figure}

We now turn to the rationalizability scores themselves. We first take up basic rationalizability ($\estar$) and EUT-rationalizability ($\etriplestar$), deferring our analysis of FOSD-rationalizability, because the distinction in between is specific to choice under uncertainty---that is, between rationalizability under the uniform prior ($\edoublestar$) and under multiple priors ($\tildeedoublestar$). In Figure \ref{fig:figure3}, we compare the distributions of $\estar$ and $\etriplestar$ across the uncertainty and risk experiments. The distributions are strikingly similar across the two domains: a two-sample Kolmogorov--Smirnov test does not reject equality of the distributions at either the 1 or the 5 percent significance levels ($p = 0.4337$ and $p = 0.2263$ for $\estar$ and $\etriplestar$, respectively). One might have conjectured that EUT-rationalizability would deteriorate as a behavioral explanation in moving from risk to uncertainty, but we find no evidence of any difference across the two domains. Given this closeness---and the analogous result for $\edoublestar$ ($p = 0.6386$) with distributions presented in Figure \ref{fig:figure4}---the move from risk to uncertainty does not by itself enlarge the scope for single-prior non-EUT models: since the distributions of both $\edoublestar$ and $\etriplestar$ are indistinguishable across the two domains, the room that probabilistic sophistication leaves over subjective EUT is of similar size under uncertainty and under risk, where \cite{dembo2026} finds it modest for most subjects. Whatever additional scope uncertainty creates must therefore come from multiple priors (more below).

Although choices under uncertainty are as rationalizable and as EUT-rationalizable as choices under risk, FOSD-rationalizability under multiple priors ($\tildeedoublestar$) accommodates the canonical models of decision making under uncertainty and allows for both ambiguity aversion and ambiguity seeking. Figure \ref{fig:figure4} compares the distributions of the FOSD-rationalizability scores $\tildeedoublestar$ and $\edoublestar$---the former allowing for multiple priors, the latter restricted to the uniform prior---where $\edoublestar$ is shown for both the uncertainty and the risk experiments; by construction, $\tildeedoublestar \geq \edoublestar$. Accordingly, the $\tildeedoublestar$ distribution in Figure \ref{fig:figure4} lies above the $\edoublestar$ distribution under uncertainty; since the vertical axis gives the fraction of subjects with a score above each value, a larger proportion of subjects clears any given threshold under multiple priors than under the uniform prior. A two-sample Kolmogorov--Smirnov test confirms that the two distributions differ significantly ($p = 0.000$). Figure \ref{fig:figure4} thus suggests that, within decision making under uncertainty, models that allow for multiple priors provide a better explanation of subjects' behavior than models restricted to the uniform prior.

\subsection{Non-Uniform Single Priors} \label{subsec:uniform}

Section \ref{subsec:aggregate} adopts the uniform prior $\mbpi = (\tfrac{1}{3}, \tfrac{1}{3}, \tfrac{1}{3})$ as the natural single-prior benchmark under uncertainty. This subsection verifies that this choice does not drive our empirical findings. We ask whether some non-uniform single prior would render choices substantially more EUT-rationalizable, and show that the uniform prior approximately maximizes EUT-rationalizability. Recall that we can test for FOSD-rationalizability and EUT-rationalizability under any single prior $\mbpi = (\pi_1, \tfrac{1}{3}, \tfrac{2}{3} - \pi_1)$, from which we obtain $\edoublestar (\pi_1)$ and $\etriplestar (\pi_1)$. The uniform prior is the natural single prior given the symmetric experimental design of \cite{ahn2014}, but we can test this hypothesis empirically across the full range of single priors.

\begin{figure}[!t]
\begin{center}
\includegraphics[scale=.5]{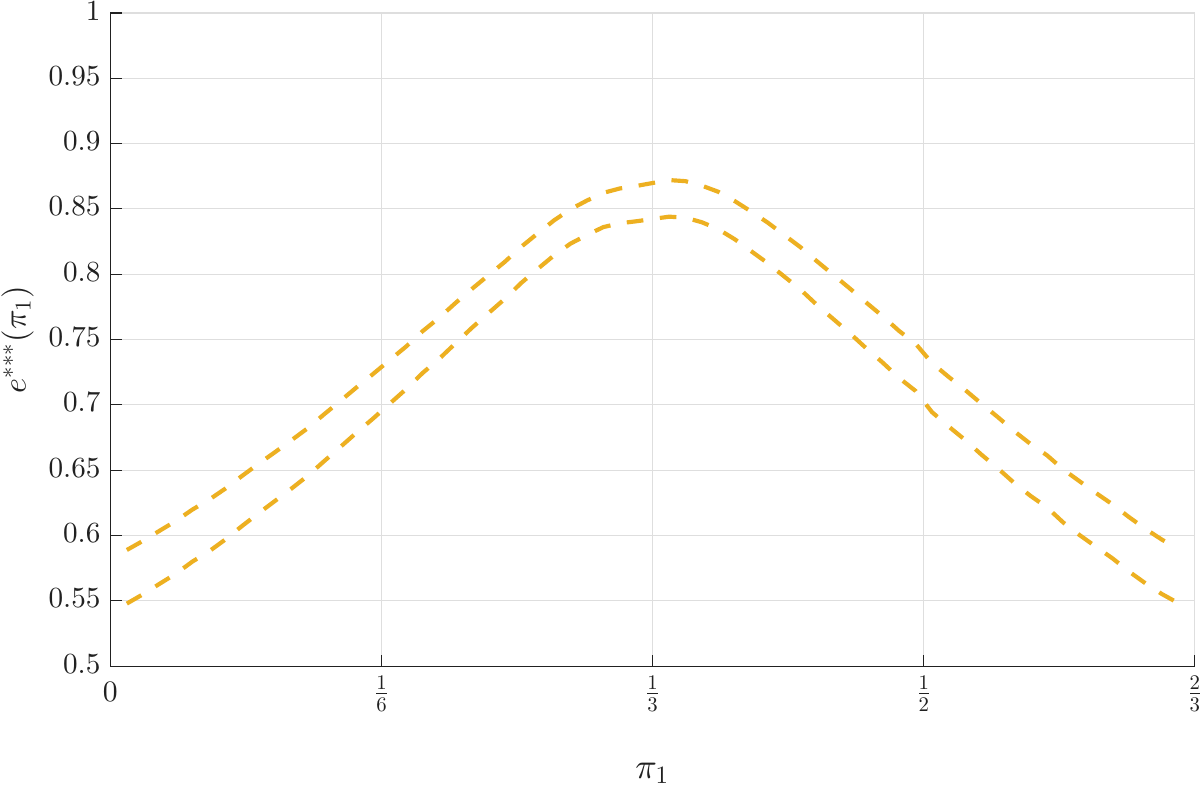}
\end{center}
\captionsetup{width=.95\linewidth}
\vspace{-.1in}
\caption{EUT-Rationalizability Scores under Single Priors}
\vspace{-.1in}
\caption*{\footnotesize{The horizontal axis shows the possible values of the probability of state 1 ($\pi_1$) and the vertical axis presents the EUT-rationalizability scores ($\etriplestar(\pi_1)$). The dashed lines indicate 95 percent confidence intervals on the mean scores across subjects.}}
\label{fig:figure5}
\end{figure}

Figure \ref{fig:figure5} plots $\etriplestar(\pi_1)$ for different values of $\pi_1$ across subjects; since $\pi_2 = \tfrac{1}{3}$ is held fixed, we let $\pi_1$ (and hence $\pi_3$) vary over the interval $(0, \tfrac{2}{3})$ along the horizontal axis, and the vertical axis depicts 95 percent confidence intervals on the mean of $\etriplestar(\pi_1)$ across subjects for each value of $\pi_1$. Figure \ref{fig:figure5} shows that $\etriplestar = \etriplestar(\tfrac{1}{3}) \approx \max_{\pi_1} \etriplestar(\pi_1)$ on average; that is, the uniform prior approximately maximizes EUT-rationalizability under a single prior in the aggregate. The same finding holds at the individual level and under FOSD-rationalizability (omitted to economize on space). The uniform prior is therefore not merely natural given the design---it is approximately the most favorable single prior---so the scope we document below for multiple-prior models cannot be attributed to an unfavorable choice of single prior.

\subsection{Individual-Level Rationalizability Scores} \label{subsec:disaggregate}

Having established that the distributions of rationalizability ($\estar$), FOSD-rationalizability ($\edoublestar$), and EUT-rationalizability ($\etriplestar$) scores are indistinguishable across the uncertainty and risk experiments, while FOSD-rationalizability under multiple priors ($\tildeedoublestar$) accounts for a larger proportion of subjects than the uniform prior, we now examine the rationalizability scores at the individual level. For each subject the scores $\estar$, $\tildeedoublestar$, and $\etriplestar$ identify the source of any departure from subjective EUT: the values of $\estar$ and $\tildeedoublestar$ reveal whether the departure stems from violations of the basic ordering axioms, from violations of first-order stochastic dominance that even multiple priors cannot accommodate, or from behavior consistent with multiple-prior non-EUT models. We first plot the scores against one another, and then compare, subject by subject, the magnitudes of the gaps between them, testing their difference using a nonparametric difference-in-differences procedure.

\begin{figure}[!t]
\begin{center}
\subfloat[$\estar$ vs. $\tildeedoublestar$]{\includegraphics[scale=.5]{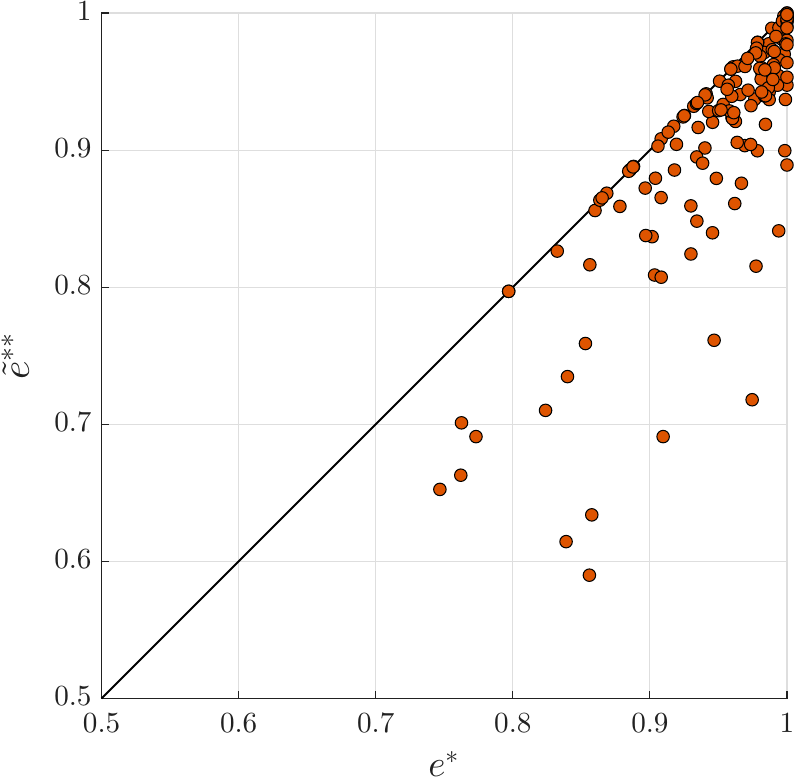}} \hspace{.5in} \subfloat[$\tildeedoublestar$ vs. $\etriplestar$]{\includegraphics[scale=.5]{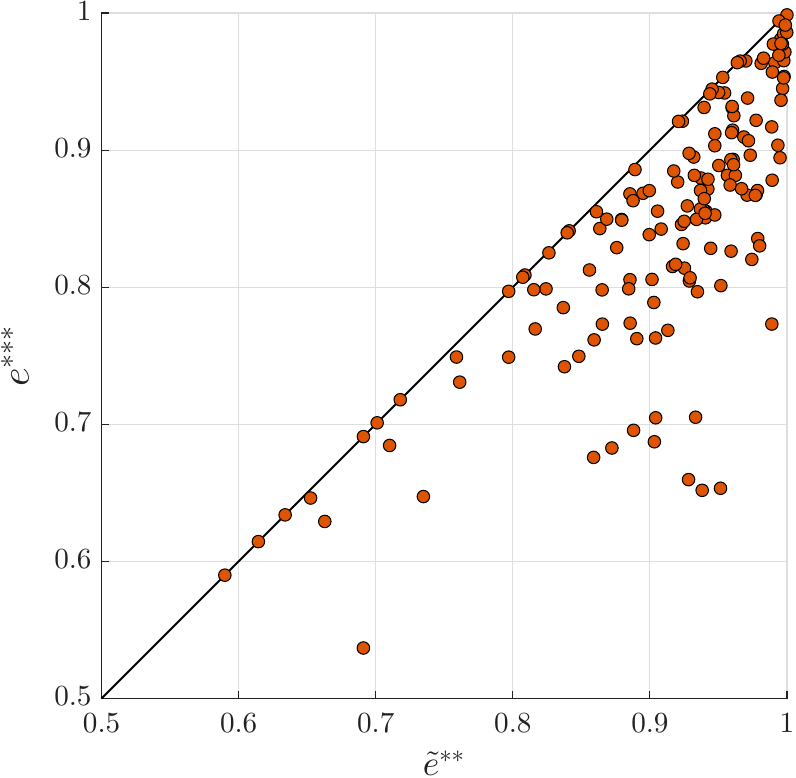}}
\end{center}
\captionsetup{width=.95\linewidth}
\vspace{-.1in}
\caption{Individual-Level Rationalizability Scores}
\vspace{-.1in}
\caption*{\footnotesize{Rationalizability scores for individual subjects. (a) $\estar$ (horizontal axis) vs. $\tildeedoublestar$ (vertical axis) and (b) $\tildeedoublestar$ (horizontal axis) vs. $\etriplestar$ (vertical axis). By definition, $\estar \geq \tildeedoublestar \geq \etriplestar$, so all points in both scatterplots must lie on or below the 45-degree line.}}
\label{fig:figure6}
\end{figure}

Figure \ref{fig:figure6}a shows a scatterplot of rationalizability ($\estar$) against FOSD-rationalizability under multiple priors ($\tildeedoublestar$), and Figure \ref{fig:figure6}b shows a scatterplot of FOSD-rationalizability under multiple priors ($\tildeedoublestar$) against EUT-rationalizability ($\etriplestar$). By definition, $\estar \geq \tildeedoublestar \geq \etriplestar$, so all points in both scatterplots lie on or below the 45-degree line. A subject who is perfectly EUT-rationalizable is at the top-right corner of both scatterplots where $1 = \estar = \tildeedoublestar = \etriplestar$; when $\etriplestar < 1$, the corresponding values of $\estar$ and $\tildeedoublestar$ identify the source of the subject's departure from subjective EUT.

Out of 154 subjects, the choices of only 20 (13.0 percent) are perfectly rationalizable ($\estar = 1$), the choices of only one (0.6 percent) are perfectly FOSD-rationalizable under multiple priors ($\tildeedoublestar = 1$), and the choices of none of the subjects are perfectly EUT-rationalizable ($\etriplestar = 1$). Furthermore, 28 subjects (18.2 percent) fall on the 45-degree line in the scatterplot of $\estar$ against $\tildeedoublestar$ (Figure \ref{fig:figure6}a); the choices of these subjects need not be perfectly rationalizable, but they are no less FOSD-rationalizable under multiple priors than they are rationalizable ($\estar = \tildeedoublestar$). Similarly, 14 subjects (9.1 percent) fall on the 45-degree line in the scatterplot of $\tildeedoublestar$ against $\etriplestar$ (Figure \ref{fig:figure6}b); the choices of these subjects are not perfectly FOSD-rationalizable under multiple priors, but they are no less EUT-rationalizable than they are FOSD-rationalizable ($\tildeedoublestar = \etriplestar$). Finally, only 1 subject (0.6 percent) falls on the 45-degree line in both scatterplots; the choices of this subject are no less EUT-rationalizable than they are rationalizable ($\estar = \tildeedoublestar = \etriplestar$).

\begin{figure}[!t]
\begin{center}
\subfloat[Full Sample ($N = 50$)]{\includegraphics[scale=.5]{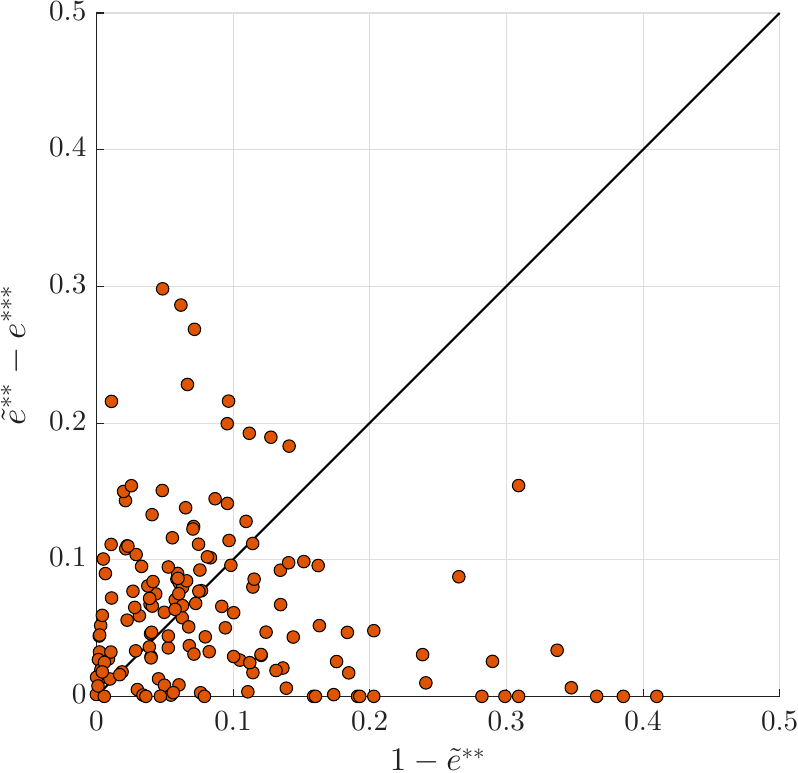}} \hspace{.5in} \subfloat[Independent Subsamples ($N = 25$)]{\includegraphics[scale=.5]{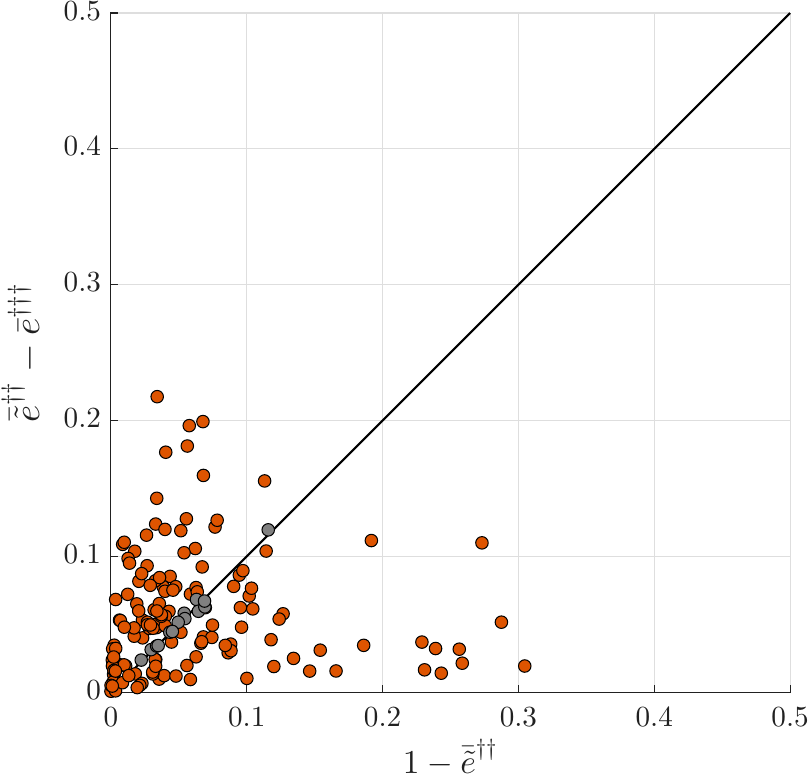}}
\end{center}
\captionsetup{width=.95\linewidth}
\vspace{-.1in}
\caption{Differences-in-Differences in Individual-Level Rationalizability Scores}
\vspace{-.1in}
\caption*{\footnotesize{Differences between perfect rationalizability and FOSD-rationalizability under multiple priors (horizontal axes) vs. differences between FOSD-rationalizability under multiple priors and EUT-rationalizability (vertical axes). (a) $(1 - \tildeedoublestar$ vs. $(\tildeedoublestar - \etriplestar)$ using the full sample of 50 observations and (b) $(1 - \bartildeedoubledagger)$ vs. $(\bartildeedoubledagger - \baretripledagger)$ using 1,000 independent subsamples of 25 observations each. In (b), subjects are depicted in red if the difference-in-difference is statistically significant at the 1 percent level and in gray if not.}}
\label{fig:figure7}
\end{figure}

Crucial to our analysis, we can also compare the magnitudes of the differences between scores. Figure \ref{fig:figure7}a shows a scatterplot of the difference between perfect rationalizability and FOSD-rationalizability under multiple priors ($1 - \tildeedoublestar$) against the difference between FOSD-rationalizability under multiple priors and EUT-rationalizability ($\tildeedoublestar - \etriplestar$). There is marked heterogeneity across subjects: of the 154 subjects, 80 (51.9 percent) fall above the 45-degree line ($1 - \tildeedoublestar < \tildeedoublestar - \etriplestar$) and 74 (48.1 percent) fall below it ($1 - \tildeedoublestar > \tildeedoublestar - \etriplestar$), so subjects are fairly evenly split on either side.

The near-even split is itself the finding. For the 80 subjects above the 45-degree line, the rationalizability that multiple-prior non-EUT models add over subjective EUT ($\tildeedoublestar - \etriplestar$) exceeds the residual departure from perfect rationalizability that no complete and transitive preference can absorb ($1 - \tildeedoublestar$); for these subjects, ambiguity-sensitive preferences account for the majority of the departure from subjective EUT, rather than that departure reflecting mere inconsistency. Pooled across subjects, the two gaps are of comparable magnitude, so models of ambiguity carry first-order---not residual---explanatory weight. This is exactly what does not arise under risk in \cite{dembo2026}, where the FOSD--EUT gap ($\edoublestar - \etriplestar$) is roughly a fifth of the perfect--FOSD gap ($1 - \edoublestar$); the even split here is direct evidence that ambiguity opens empirical space for non-EUT models that the risk setting does not.

The rich data of \cite{ahn2014} also allow us to compare, for each individual subject, the gap between perfect rationalizability and FOSD-rationalizability under multiple priors with the gap between FOSD-rationalizability under multiple priors and EUT-rationalizability, using a difference-in-differences approach. The test is purely nonparametric, in that no functional form assumptions are imposed on subjects' underlying preferences or on the error structure. To construct a test statistic and sampling distribution for each subject, we repeatedly subsample half of the subject's data, drawing 1,000 subsamples of 25 observations each; see \cite{dembo2026} for a description of the test, including the subsampling and bootstrapping procedures. For each subsample, we calculate FOSD-rationalizability under multiple priors and EUT-rationalizability scores, denoted by $\tildeedoubledagger$ and $\etripledagger$, and the resulting difference-in-differences $(1 - \tildeedoubledagger) - (\tildeedoubledagger - \etripledagger)$. We use $\dagger$ rather than $\ast$ to denote rationalizability scores based on 25 rather than 50 observations.

Figure \ref{fig:figure7}b displays the result of this test on the independent subsamples, coloring each subject red when the difference-in-differences is statistically significant at the 1 percent level and gray otherwise. The difference-in-differences fails to reach significance for only a handful of subjects---14 of the 154 (9.1 percent)---so for the remaining the test reliably places the subject on one side of the 45-degree line. This matters because subsampling uses only half of a subject's data, resulting in fewer observed violations. That the test nonetheless rejects equality of the two gaps for nearly every subject establishes that each subject's position relative to the 45-degree line is a genuine, individual-level feature of behavior, not an artifact of how the scores are constructed or of sampling noise.

\subsection{Power} \label{subsec:power}

The wedge between FOSD-rationalizability under multiple priors and under the uniform prior is the empirical space in which multiple-prior non-EUT models can improve on subjective EUT. This subsection asks whether that wedge reflects genuine, systematically ambiguity-averse behavior or is merely a mechanical consequence of the additional freedom that multiple priors afford. In the same spirit as \cite{dembo2026}, we answer this by comparing the wedge of our actual subjects with the wedge of simulated subjects whose departures from uniform-prior FOSD-rationalizability are unstructured.

\begin{figure}[!t]
\begin{center}
\includegraphics[scale=.5]{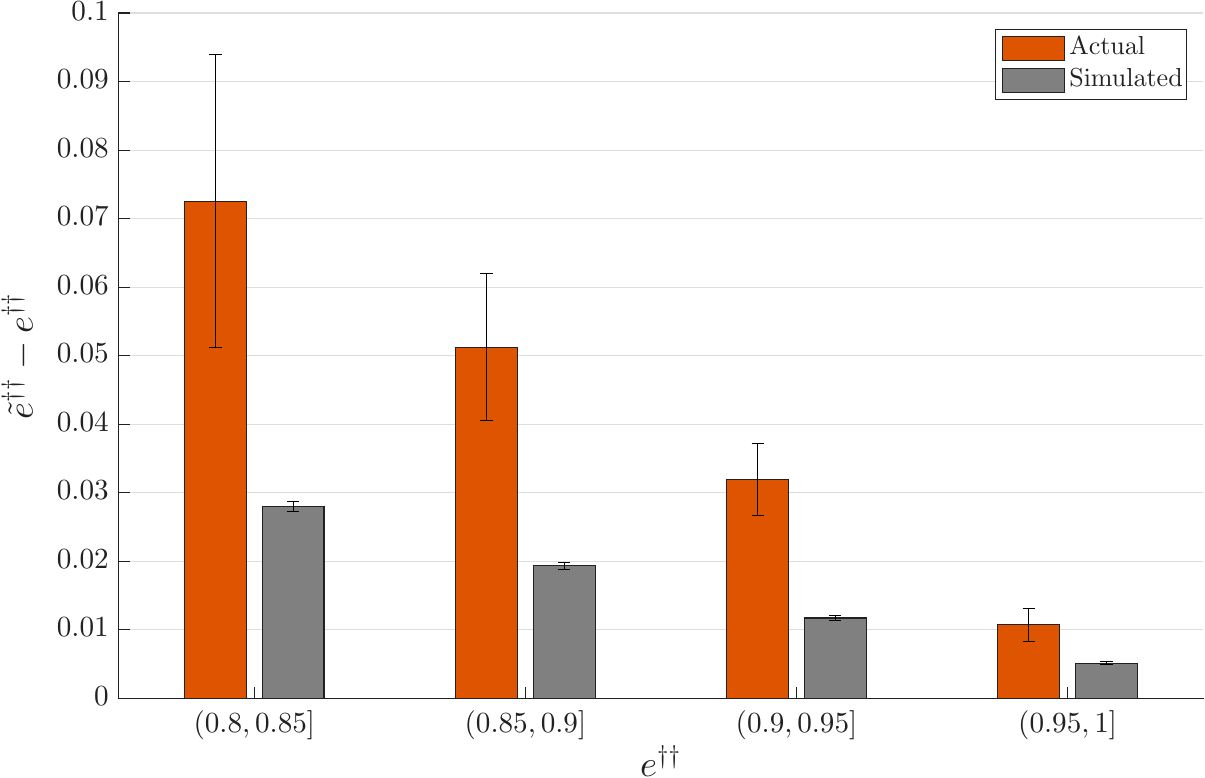}
\end{center}
\captionsetup{width=.95\linewidth}
\vspace{-.1in}
\caption{Power of Testing FOSD-Rationalizability}
\vspace{-.1in}
\caption*{\footnotesize{The horizontal axis shows different score value bins for FOSD-rationalizability under the uniform prior ($\edoubledagger$). The vertical axis represents the gap between FOSD-rationalizability under multiple priors versus under the uniform prior ($\tildeedoubledagger - \edoubledagger$). The braces represent 95 percent confidence intervals.}}
\label{fig:figure8}
\end{figure}

Following \cite{dembo2026} and in the spirit of \cite{bronars1987}, the simulated subjects choose uniformly at random on each budget set, conditional on attaining a given level of FOSD-rationalizability under the uniform prior; that is, their choices are FOSD-rationalizable under the uniform prior up to a controlled degree of error, as measured by $\edoubledagger$. Each simulated subject makes choices from budget sets drawn from the same distribution as the actual subjects, and for each we compute the multiple-prior FOSD-rationalizability score $\tildeedoubledagger$ and the wedge $\tildeedoubledagger - \edoubledagger$. To compare the two groups at a common number of decisions, the actual subjects' scores are averaged over 1,000 subsamples of 25 choices (and thus we use $\dagger$ rather than $\ast$ to denote rationalizability scores). See \cite{dembo2026} for precise details of the simulation. Figure \ref{fig:figure8} reports the results. The horizontal axis groups subjects into bins by their uniform-prior FOSD-rationalizability score $\edoubledagger$, and the vertical axis shows, within each bin, the mean wedge $\tildeedoubledagger - \edoubledagger$ for the actual and the simulated subjects.

Across the range of uniform-prior FOSD-rationalizability scores, the wedge is larger for the actual subjects than for the simulated subjects. Because the two groups share the same uniform-prior FOSD score $\edoubledagger$ within each bin, this gap is due entirely to the actual subjects attaining higher multiple-prior FOSD-rationalizability $\tildeedoubledagger$: their departures from uniform-prior FOSD-rationalizability are systematically of the kind that multiple priors can accommodate, whereas the simulated subjects' unstructured departures are not. The behavior of many actual subjects therefore cannot be explained as uniform-prior FOSD-rationalizability subject to error---the pattern characteristic of the simulated subjects---because such error does not open a wedge between multiple-prior and uniform-prior FOSD-rationalizability. The larger wedge among the actual subjects instead indicates genuine ambiguity aversion of the kind generated by the kinked and smooth specifications, which are FOSD-rationalizable under multiple priors but not under the uniform prior.

\subsection{Kinked Specification} \label{subsec:kinked}

To complete the analysis, we also test for rationalizability by the kinked specification, interpreted through the lens of the $\alpha$-MEU model \citep{ghirardato2004} as presented in Section \ref{sec:models}. As described in Section \ref{subsec:scores}, for each value of the ambiguity parameter $\alpha \in (0,1)$, we test whether the data can be rationalized by a utility function of the form in equation (\ref{eq:kinked}) using the GRID method of \cite{polisson2020}, making no assumptions about the structure of the Bernoulli index $u$; this yields the score $e^{\kappa}(\alpha)$ for each $\alpha$, and $e^{\kappa} = \max_{\alpha} e^{\kappa}(\alpha)$, with $\etriplestar \leq e^{\kappa} \leq \tildeedoublestar$, and no ordering relative to $\edoublestar$. We do not conduct an analogous test of the smooth specification \citep{klibanoff2005}, for the reasons given in Sections \ref{sec:models} and \ref{sec:analysis}.\footnote{Neither do we perform a test of the generalized kinked specification \citep{segal1987,segal1990}. This analysis is also feasible in principle but time-consuming due to its computational complexity. In any case, the score of this specification can be no higher than $\tildeedoublestar$: like the other models in the multiple-prior class, the generalized kinked specification obeys FOSD and preserves the symmetry between the uncertain states, and $\tildeedoublestar$ is the upper bound on the performance of any model with these properties.}

The illustrative subjects of Section \ref{sec:subjects} already indicate the range of outcomes. Where subjective EUT leaves no room for improvement, $e^{\kappa}$ is virtually indistinguishable from $\etriplestar$; where it does leave room, the kinked specification captures anywhere from about half to all of the scope available to the multiple-prior class. We now turn to the full sample.

\begin{figure}[!t]
\begin{center}
\subfloat[$e^{\kappa}(\alpha)$]{\includegraphics[scale=.5]{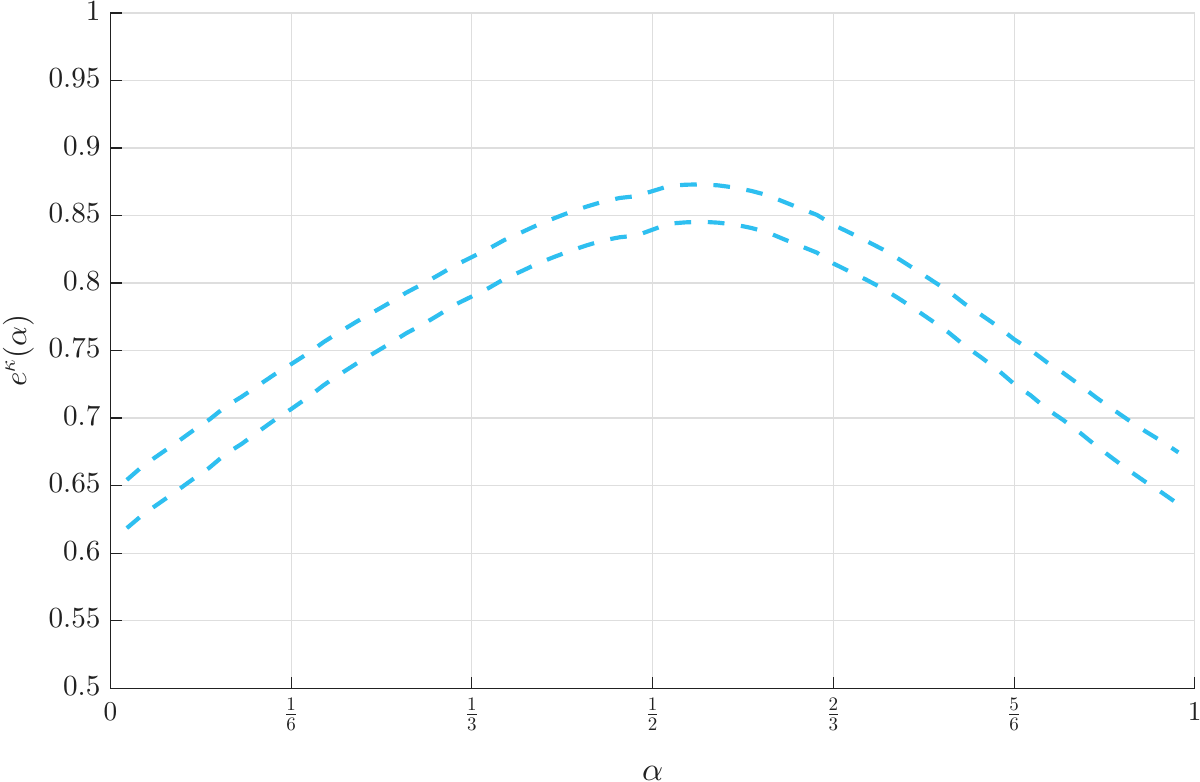}} \\ \subfloat[$F(\alpha)$]{\includegraphics[scale=.5]{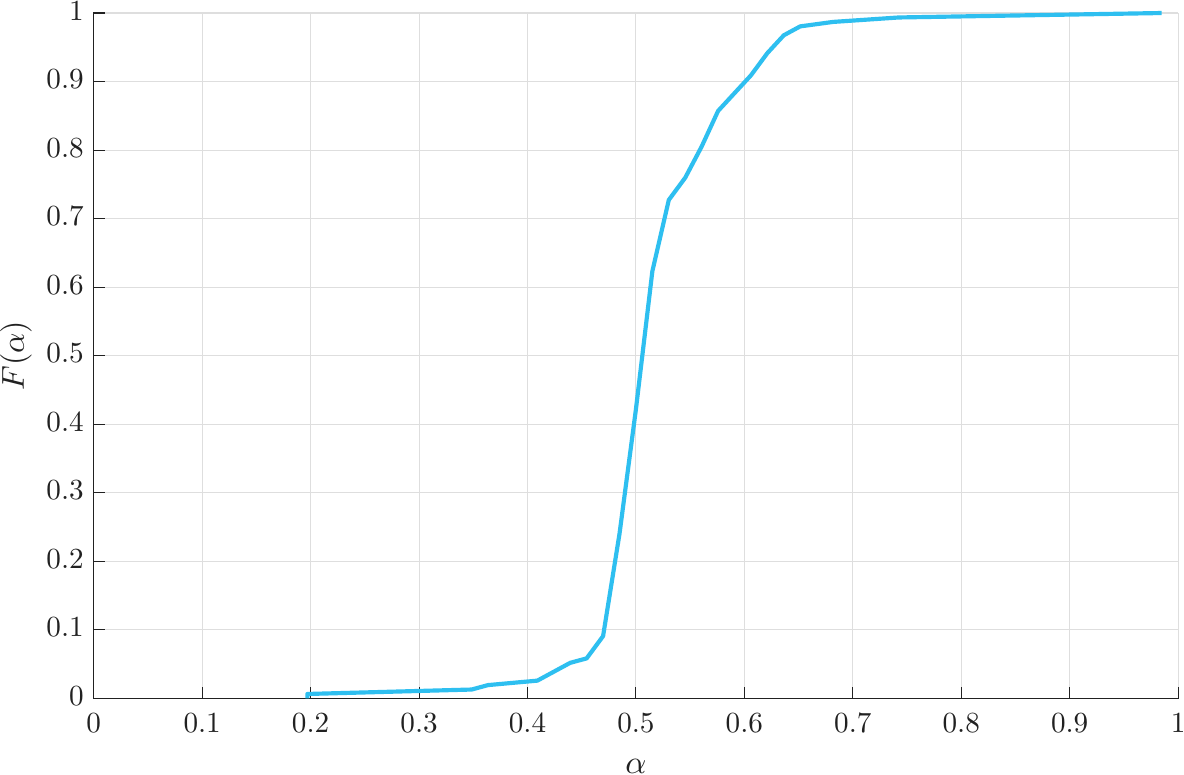}}
\end{center}
\captionsetup{width=.95\linewidth}
\vspace{-.1in}
\caption{Kinked-Rationalizability}
\vspace{-.1in}
\caption*{\footnotesize{(a) The horizontal axis shows the possible values of the ambiguity parameter ($\alpha$) and the vertical axis presents the kinked-specification rationalizability scores ($e^{\kappa}(\alpha)$). The dashed lines indicate 95 percent confidence intervals on the mean scores across subjects. (b) Cumulative distribution function $F(\alpha)$ of the score-maximizing values of $\alpha$ (when multiple values of $\alpha$ maximize $e^{\kappa}(\alpha)$, the value closest to $\alpha = \tfrac{1}{2}$ is selected).}}
\label{fig:figure9}
\end{figure}

Figure \ref{fig:figure9}a examines the kinked-specification rationalizability scores across the full range of the ambiguity parameter, just as Figure \ref{fig:figure5} does for the EUT-rationalizability scores across the full range of single priors. The horizontal axis shows the value of $\alpha$, and the vertical axis depicts 95 percent confidence intervals on the mean of $e^{\kappa}(\alpha)$ across subjects for each value of $\alpha$. Note that at $\alpha = \tfrac{1}{2}$, the kinked specification reduces to subjective EUT under the uniform prior, so $e^{\kappa}(\tfrac{1}{2}) = \etriplestar$ by construction. The mean scores are lowest around 0.62 to 0.65, at the extremes of $\alpha$ near zero (extreme ambiguity seeking, the maxmax case) and near one (extreme ambiguity aversion, the maxmin case), and rise steadily toward a peak slightly to the right of $\alpha = \tfrac{1}{2}$. Just as Figure \ref{fig:figure5} shows that the uniform prior approximately maximizes EUT-rationalizability, Figure \ref{fig:figure9}a shows that the best-fitting ambiguity parameter lies close to (but slightly above) the subjective EUT benchmark of $\alpha = \tfrac{1}{2}$, consistent with overall mild ambiguity aversion on average.

Figure \ref{fig:figure9}b turns from the aggregate to the individual level and shows the distribution of the value of $\alpha$ that maximizes $e^{\kappa}(\alpha)$ subject by subject. When there are multiple values of $\alpha$ that maximize $e^{\kappa}(\alpha)$ for a given subject, we select the maximizing value closest to $\alpha = \tfrac{1}{2}$, that is, closest to the parsimonious special case of subjective EUT. The distribution is tightly concentrated around $\alpha = \tfrac{1}{2}$: there is little mass below 0.45, the distribution rises steeply between roughly 0.45 and 0.6, and nearly all subjects lie below 0.65. The median is slightly above $\tfrac{1}{2}$, and the mass is tilted to the right of $\tfrac{1}{2}$, consistent with the overall mild ambiguity aversion evident in Figure \ref{fig:figure9}a.

\begin{figure}[!t]
\begin{center}
\subfloat[$\tildeedoublestar$, $e^{\kappa}$, $\edoublestar$, $\etriplestar$]{\includegraphics[scale=.37]{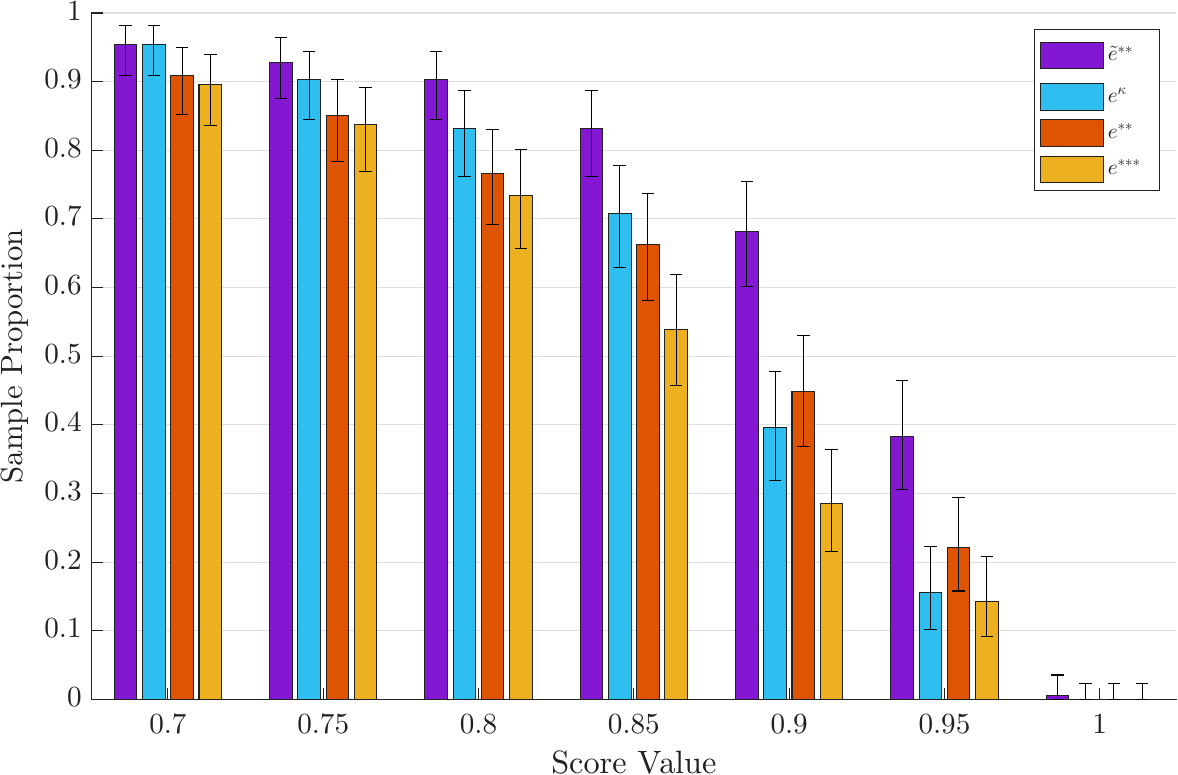}} \hspace{.5in} \subfloat[$\tildeedoublestar$, $\max \, \{e^{\kappa}, \edoublestar\}$, $\etriplestar$]{\includegraphics[scale=.37]{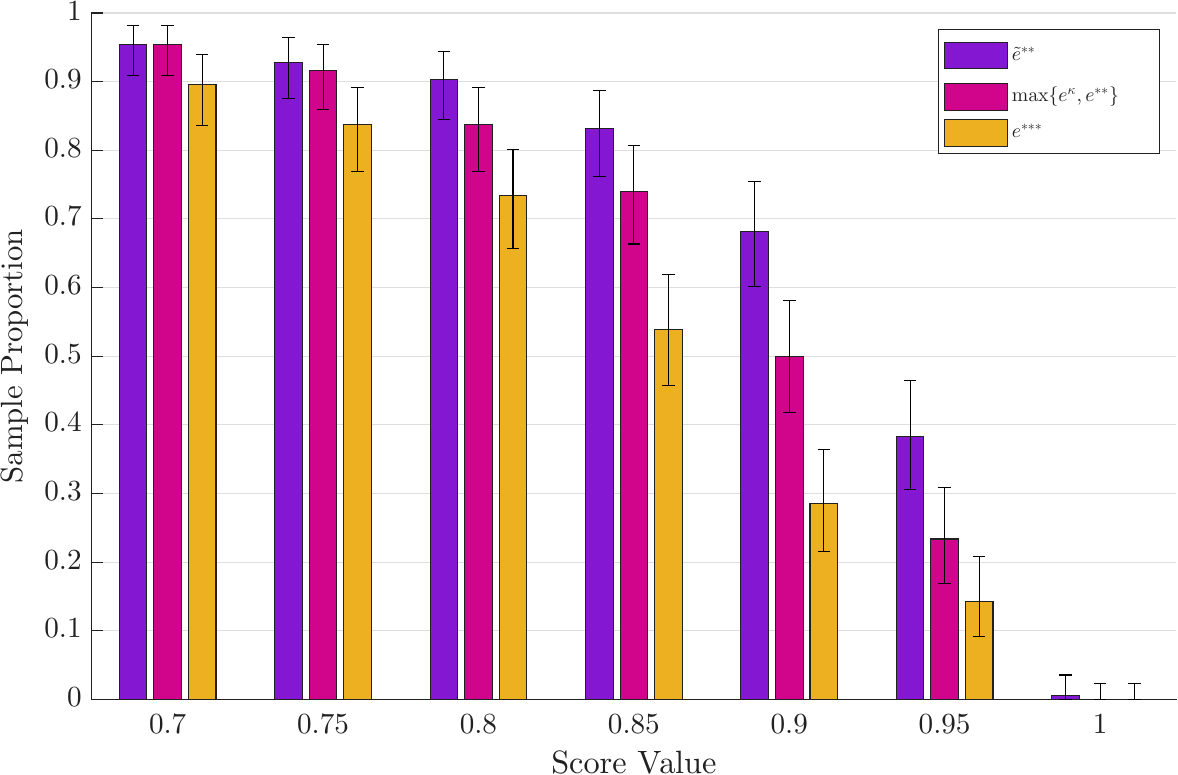}}
\end{center}
\captionsetup{width=.95\linewidth}
\vspace{-.1in}
\caption{Distributions of Rationalizability Scores}
\vspace{-.1in}
\caption*{\footnotesize{The horizontal axis shows the score values and the vertical axis presents the fraction of subjects with scores above each value for (a) $\tildeedoublestar$, $e^{\kappa}$, $\edoublestar$, and $\etriplestar$; and (b) $\tildeedoublestar$, $\max \, \{e^{\kappa}, \edoublestar\}$, and $\etriplestar$. The braces indicate 95 percent confidence intervals on these proportions.}}
\label{fig:figure10}
\end{figure}

Figure \ref{fig:figure10}a presents the distributions of the scores in the format used throughout the paper: the horizontal axis shows score values, and the vertical axis presents the fraction of subjects with scores above each value for $\tildeedoublestar$, $e^{\kappa}$, $\edoublestar$, and $\etriplestar$. As with the other rationalizability scores, the $e^{\kappa}$ scores are high (95.5 percent of subjects have $e^{\kappa}$ above 0.7, and 70.8 percent above 0.85), so most of the action in the data is in the rightmost bars. As in Figure \ref{fig:figure1}, Figure \ref{fig:figure10}a reveals a layered pattern in which different gaps are more pronounced at different points of the score distribution. At the 0.7 threshold the $e^{\kappa}$ and $\tildeedoublestar$ bars coincide (both 95.5 percent), and at 0.8 they remain close (83.1 and 90.3 percent); at both thresholds $e^{\kappa}$ lies above $\edoublestar$ (90.9 and 76.6 percent). At moderate consistency levels, the kinked specification accommodates nearly as many subjects as the entire multiple-prior class and more than probabilistic sophistication. Around 0.85 the gap between $\tildeedoublestar$ (83.1 percent) and $e^{\kappa}$ (70.8 percent) becomes prominent, while $e^{\kappa}$ remains above $\edoublestar$ (66.2 percent). At the 0.9 and 0.95 thresholds, however, the ordering of the middle bars flips (44.8 versus 39.6 percent at 0.9, and 22.1 versus 15.6 percent at 0.95), which is possible precisely because $e^{\kappa}$ and $\edoublestar$ are not nested. Thus, at the highest consistency levels, models of probabilistic sophistication account for more subjects than the kinked specification, whereas the kinked specification accounts for more subjects across the middle of the distribution.

Figure \ref{fig:figure10}b replaces the two middle scores with $\max \, \{e^{\kappa}, \edoublestar\}$ for each subject, the better of the two non-nested, non-EUT directions subject by subject, presented alongside $\tildeedoublestar$ and $\etriplestar$. The fraction of subjects with $\max \, \{e^{\kappa}, \edoublestar\}$ above 0.85 is 74.0 percent, compared with 83.1 percent for $\tildeedoublestar$ and 53.9 percent for $\etriplestar$; at 0.9 the corresponding fractions are 50.0, 68.2, and 28.6 percent, and at 0.95 they are 23.4, 38.3, and 14.3 percent. The gap between the $\max \, \{e^{\kappa}, \edoublestar\}$ bars and the $\etriplestar$ bars measures the improvement over EUT delivered by the better of the kinked specification and probabilistic sophistication, while the remaining gap up to $\tildeedoublestar$ measures the part of the multiple-prior scope that neither captures.

\begin{figure}[!t]
\begin{center}
\subfloat[$\edoublestar$ vs. $e^{\kappa}$]{\includegraphics[scale=.5]{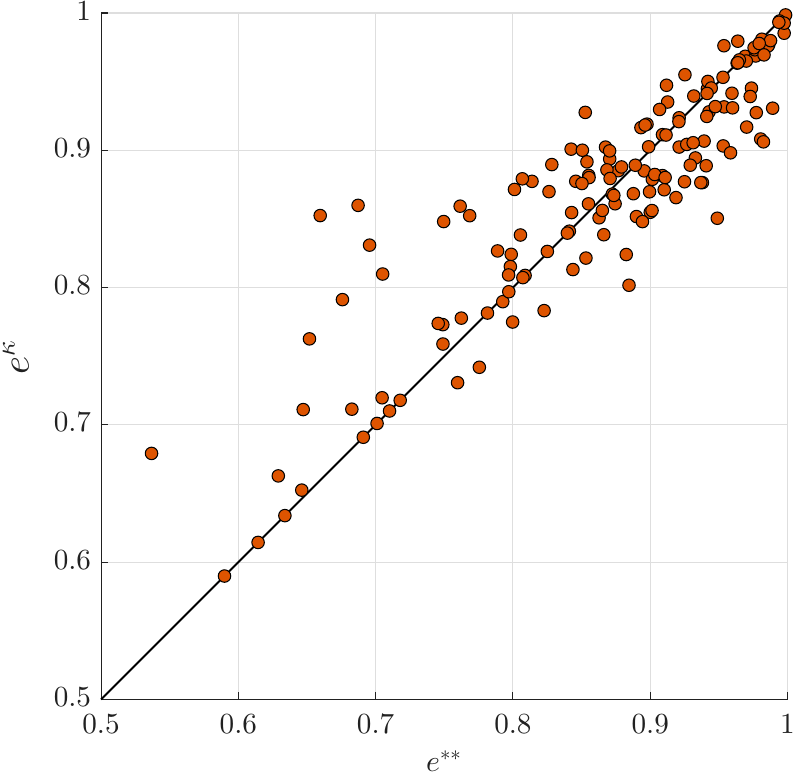}} \hspace{.5in} \subfloat[$\max \, \{e^{\kappa}, \edoublestar\}$ vs. $\etriplestar$]{\includegraphics[scale=.5]{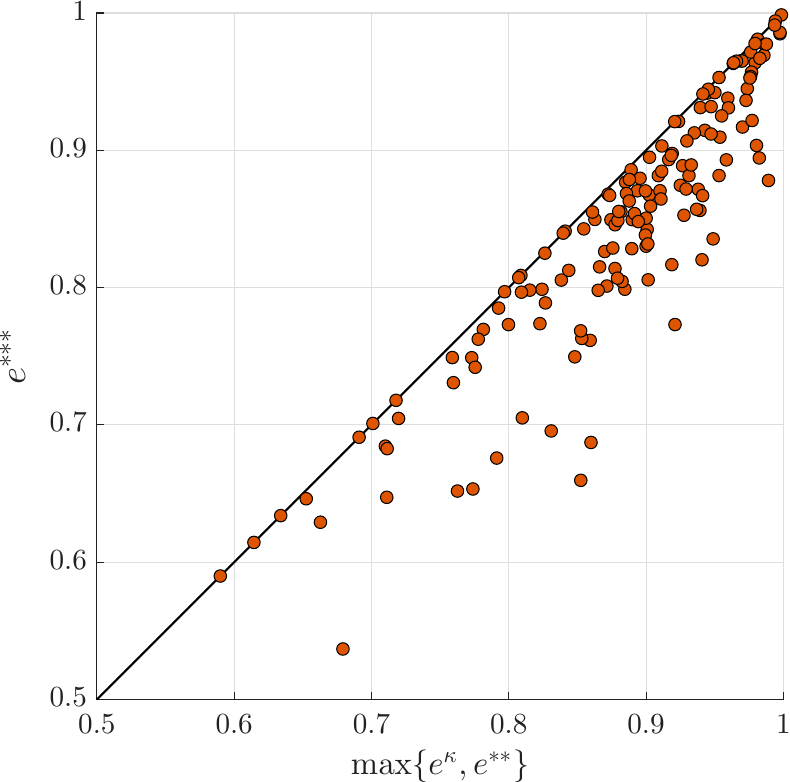}}
\end{center}
\captionsetup{width=.95\linewidth}
\vspace{-.1in}
\caption{Individual-Level Rationalizability Scores}
\vspace{-.1in}
\caption*{\footnotesize{Rationalizability scores for individual subjects. (a) $\edoublestar$ (horizontal axis) vs. $e^{\kappa}$ (vertical axis) and (b) $\max \, \{e^{\kappa}, \edoublestar\}$ (horizontal axis) vs. $\etriplestar$ (vertical axis). By definition, $\max \, \{e^{\kappa}, \edoublestar\} \geq \etriplestar$, so all points in (b) must lie on or below the 45-degree line; by contrast, since $e^{\kappa}$ and $\edoublestar$ are not ordered by construction, points in (a) can lie above or below the 45-degree line.}}
\label{fig:figure11}
\end{figure}

Finally, Figure \ref{fig:figure11}a shows a scatterplot of the individual-level scores $\edoublestar$ (horizontal axis) against $e^{\kappa}$ (vertical axis). Because the two scores are not nested, observations lie on both sides of the 45-degree line. Of the 154 subjects, 66 (42.9 percent) lie above the 45-degree line ($\edoublestar < e^{\kappa}$), so their choices are better explained by the kinked specification; 69 (44.8 percent) lie below the line ($\edoublestar > e^{\kappa}$), so their choices are better explained by probabilistic sophistication; and 19 (12.3 percent) lie on the line. The sample thus splits almost evenly between the two non-EUT directions, underscoring once more the heterogeneity across subjects that runs through our individual-level results: neither the kinked specification nor probabilistic sophistication dominates the other as a description of behavior.

Figure \ref{fig:figure11}b shows a scatterplot of $\max \, \{e^{\kappa}, \edoublestar\}$ (horizontal axis) against $\etriplestar$ (vertical axis). Since both $e^{\kappa}$ and $\edoublestar$ are weakly greater than $\etriplestar$, all observations must lie on or below the 45-degree line. For observations on the line, non-EUT models, whether the kinked specification or probabilistic sophistication, do not outperform subjective EUT at all: 15 subjects (9.7 percent) fall on the line, while the remaining 139 subjects (90.3 percent) fall below it. More importantly, the improvement is economically meaningful for many subjects: $\max \, \{e^{\kappa}, \edoublestar\} - \etriplestar$ exceeds 0.025 for 82 subjects (53.2 percent) and exceeds 0.05 for 44 subjects (28.6 percent). Recalling the money-metric interpretation of the CCEI, for over half of the subjects the better of the two non-EUT models reduces the fraction of the budget left on the table relative to EUT by more than 2.5 percent of wealth, and for more than a quarter of the subjects by more than 5 percent, a real improvement for those subjects.

\section{Concluding Remarks} \label{sec:conclusion}

This paper has carried the nonparametric revealed preference methodology of \cite{dembo2026} from choice under risk to choice under uncertainty, applying it to the portfolio choice data of \cite{ahn2014}. As in its companion, the analysis is both nonparametric---resting only on revealed preference relations---and comprehensive---testing the full set of axioms underpinning subjective EUT and its leading non-EUT alternatives. The central construct is a nested chain of CCEI-type scores, $1 \geq \estar \geq \tildeedoublestar \geq \edoublestar \geq \etriplestar > 0$, evaluating each subject's data against basic rationalizability, FOSD-rationalizability, and EUT-rationalizability in turn. The score $\tildeedoublestar$, FOSD-rationalizability under multiple priors, has no counterpart in the risk setting; it bounds the explanatory reach of the entire multiple-prior class without committing to a particular specification.

Three findings stand out. First, choices under uncertainty are as rationalizable and as EUT-rationalizable as choices under risk: the distributions of $\estar$ and $\etriplestar$ are statistically indistinguishable from their risk-experiment counterparts. Whatever additional cognitive burden subjective uncertainty imposes, it does not erode the consistency of behavior with utility maximization, or with expected-utility maximization, relative to objective risk. Furthermore, in both the risk and uncertainty settings, departures from basic rationalizability, where they arise, frequently involve WARP violations and not just GARP-not-WARP violations, therefore reflecting incompleteness and not just nontransitivity.

Second, and in sharp contrast to the risk setting, the data leave considerable room for models of ambiguity aversion. The gap between multiple-prior and uniform-prior FOSD-rationalizability ($\tildeedoublestar - \edoublestar$) is the dominant feature across much of the score distribution, whereas the analogous uniform-prior gap that survives under risk is modest for the vast majority of subjects. This wedge is what opens empirical space for the non-EUT models that accommodate ambiguity aversion, and it is precisely the layer the risk experiment lacks. An individual-level difference-in-differences test, applied to repeated subsamples of each subject's data, confirms that these gaps between scores are reliable features of behavior rather than artifacts of the score construction, and a synthetic-data power analysis shows that the gap among actual subjects exceeds that which arises among simulated subjects, supporting its reading as genuine ambiguity aversion.

Third, a test of the kinked specification, interpreted through the lens of $\alpha$-MEU, shows that a parsimonious one-parameter generalization of subjective EUT and probabilistic sophistication are complementary explanations that together capture much, though not all, of this scope. At moderate levels of consistency $e^{\kappa}$ lies close to the multiple-prior upper bound $\tildeedoublestar$ and accommodates more subjects than probabilistic sophistication, whereas at the highest levels of consistency the ordering reverses, and at the individual level the sample splits almost evenly between subjects better explained by each. Taking the better of the two for each subject, $\max \, \{e^{\kappa}, \edoublestar\}$, raises the fraction of subjects with scores above 0.9 from 28.6 percent under subjective EUT to 50.0 percent, against 68.2 percent for $\tildeedoublestar$, and the improvement over subjective EUT exceeds 2.5 percent of wealth for over half of the subjects.

Our results also bear on the practice of structural estimation. When choice data depart materially from rationalizability or FOSD-rationalizability, fitting a parametric utility family risks recovering parameters of a preference that does not describe the subject, with consequences for positive prediction and welfare analysis---the concern pressed by \cite{halevy2018}. By measuring how far the data lie from each model nonparametrically and at the individual level, before any functional form is imposed, the methodology offers a disciplined way to gauge when a given class of preferences can be taken to the data, and when it cannot. Where \cite{savage1954} reduced uncertainty to risk, the revealed preference record studied here shows that \citeauthor{ellsberg1961}'s (\citeyear{ellsberg1961}) intuition leaves a measurable footprint---one that subjective expected utility cannot absorb but multiple-prior models can.

Finally, the departures from basic rationalizability documented in our paper are due to failures of transitivity, completeness, or both, and the literature offers models of each. On nontransitivity, the state-based regret apparatus of \cite{bell1982}, \cite{fishburn1982}, and \cite{loomes1982} was developed for risk but transfers to the Savage setting, and \cite{fishburn1984,fishburn1989}, \cite{sugden1993}, \cite{quiggin1994}, and \cite{nakamura1998} supply axiomatic foundations for nontransitive subjective expected utility---a single-prior representation that relaxes transitivity alone (see also \cite{diecidue2017}). On incompleteness, dropping completeness yields multi-utility representations under risk \citep{dubra2004} and multi-prior, multi-utility representations under uncertainty, in which non-comparability is traceable to indecisiveness in beliefs, in tastes, or in both \citep{ok2012,galaabaatar2013,riella2015,karni2026}.
 
In the revealed preference tradition, \cite{eliaz2006} weakens WARP to a non-inferiority axiom that admits rationalization by an incomplete preference. \cite{karni2023} brings the apparatus into a portfolio problem over Arrow securities---the theoretical analog of our environment---where greater incompleteness of beliefs or tastes enlarges the set of undominated positions, and constructs measures of comparative incompleteness that suggest a route to quantifying, subject by subject, what our scores detect. That both departures appear together in our data points to a gap the theory has yet to close. The multi-utility tradition relaxes completeness while keeping transitivity, and the regret and nontransitive-utility tradition relaxes transitivity while keeping completeness; a model that does both---incomplete preferences whose strict part may cycle---is coherent in principle, as \cite{mandler2005} and \cite{nishimura2016} establish, but it has neither been cast in the mixture-space, independence-based form the canonical risk and uncertainty models share nor confronted with data. Casting such a model and testing it nonparametrically is an important direction for theoretical and experimental research on choice under uncertainty.

\linespread{1.0} \selectfont \small

\bibliography{rpbib}

\appendix

\linespread{1.5} \selectfont \normalsize

\section{Introductory Example}

In the example presented in the Introduction, the allocation $\mbx^1 = (x_M, x_H, x_L)$ is chosen at the prices $\mbp^1 = (p_L, p_M, p_H)$, and the allocation $\mbx^2 = (x_L, x_H, x_M)$ is chosen at the prices $\mbp^2 = (p_H, p_M, p_L)$, where $p_H > p_M > p_L$ and $x_H > x_M > x_L$. These choices are not compatible with the maximization of a utility function that is FOSD-increasing under any single prior $\mbpi = (\pi_1, \tfrac{1}{3}, \pi_3)$ satisfying $\pi_1 + \pi_3 = \tfrac{2}{3}$. Indeed, suppose the choices are obtained by maximizing such a utility function. Then in the first budget, $\pi_1 < \tfrac{1}{3}$; otherwise, $(x_H + \epsilon, x_M + \epsilon, x_L)$ for $\epsilon > 0$ sufficiently small costs less than $\mbx^1$ at $\mbp^1$ and delivers higher utility. Since the example is symmetric, an analogous argument can be made for the second budget, so $\pi_3 < \tfrac{1}{3}$. But both inequalities cannot hold simultaneously, since $\pi_1 + \pi_3 = \tfrac{2}{3}$.

However, $\mbp^1 \cdot \mbx^2 > \mbp^1 \cdot \mbx^1$ and $\mbp^2 \cdot \mbx^1 > \mbp^2 \cdot \mbx^2$, so the data satisfy GARP and are thus consistent with the maximization of a well-behaved utility function. In fact, we can be more specific. We make the following claim:

\begin{claim}
Provided $p_M \leq (p_L + p_H) / 2$ and $x_L > 0$, the dataset $\{(\mbp^1, \mbx^1), (\mbp^2, \mbx^2)\}$ is consistent with MEU \citep{gilboa1989}.
\end{claim}

\begin{proof}
Given a Bernoulli index $u$ and a probability distribution $\mbpi$, let $\mathbb{E}_{\mbpi} u(\mbx)$ denote the expected utility of $\mbx$ given the distribution $\mbpi$. Consider a MEU model where
\[
U(\mbx) = \min_{\mbpi \in \Pi} \, \mathbb{E}_{\mbpi} u(\mbx),
\]
where $\Pi$ is a set of probability distributions. The function $U$ is concave so long as $u$ is concave, because it is the minimum of a set of concave functions. Suppose that
\[
\Pi = \left\{ \left( \frac{2}{3} \, a_1, \frac{1}{3}, \frac{2}{3} \, a_3 \right) : a_1 + a_3 = 1 \, \, \text{and} \, \, a_1, a_3 \geq \underline{a} \right \}
\]
for some $\underline{a} \in [0, \tfrac{1}{2}]$; this represents a set of probability distributions that are symmetric between states 1 and 3. Then we may re-write $U$ as
\[
U(\mbx) = \frac{2}{3} \, (1 - \underline{a}) \, u(\min \{x_1, x_3\}) + \frac{1}{3} \, u(x_2) + \frac{2}{3} \, \underline{a} \, u(\max \{x_1, x_3\})
\]
so long as $u$ is strictly increasing.

Suppose that the dataset $\{(\mbp^1, \mbx^1), (\mbp^2, \mbx^2)\}$ can be rationalized by MEU. Then the allocation $\mbx^1 = (x_M, x_H, x_L) \gg \mathbf{0}$ has utility
\[
U(\mbx^1) = \frac{2}{3} \, (1 - \underline{a}) \, u(x_L) + \frac{1}{3} \, u(x_H) + \frac{2}{3} \, \underline{a} \, u(x_M).
\]
Assuming that $u$ is differentiable, with $u' > 0$ and $u'' \leq 0$, this allocation is optimal at $\mbp^1 = (p_L, p_M, p_H)$ whenever the following first-order condition holds:
\[
\frac{\tfrac{2}{3} \, \underline{a} \, u'(x_M)}{p_L} = \frac{\tfrac{1}{3} \, u'(x_H)}{p_M} = \frac{\tfrac{2}{3} \, (1 - \underline{a}) \, u'(x_L)}{p_H}.
\]
This equation admits a solution, in the sense that there is a Bernoulli index $u$, with $u' > 0$ and $u'' \leq 0$, that satisfies this equation, if and only if
\[
\frac{2 (1 - \underline{a})}{p_H} \leq \frac{2 \underline{a}}{p_L} \leq \frac{1}{p_M},
\]
since $u'(x_L) \geq u'(x_M) \geq u'(x_H)$. It is easy to check that there is $\underline{a}$ that solves this inequality if and only if $p_M \leq (p_L + p_H) / 2$; if this condition holds we can choose $\underline{a} = p_L / 2p_M$.
\end{proof}

\section{Kinked Specification}

In this section, we provide proofs for the claims made in Section 4.2.

\begin{claim}
For any $\bar{\alpha} \in [0, 1]$ and $\Pi = \left\{ \left( a_1, \tfrac{1}{3}, a_3 \right) : a_1 + a_3 = \tfrac{2}{3} \, \, \text{and} \, \, a_1, a_3 \geq \underline{a} \right\}$, $\underline{a} \in (0, \tfrac{1}{3}]$,
\begin{align*}
U(\mbx) & = \bar{\alpha} \, \min_{\mbpi \in \Pi} \, \mathbb{E}_{\mbpi} u(\mbx) + (1 - \bar{\alpha}) \, \max_{\mbpi \in \Pi} \, \mathbb{E}_{\mbpi} u(\mbx) \\
& = \frac{2}{3} \, \alpha \, u(x_{\min}) + \frac{1}{3} \, u(x_2) + \frac{2}{3} \, (1 - \alpha) \, u(x_{\max})
\end{align*}
for some $\alpha \in (0,1)$.
\end{claim}

\begin{proof}
With no loss of generality, suppose $x_1 \geq x_3$. Then
\begin{align*}
U(\mbx) & = \bar{\alpha} \, \min_{\mbpi \in \Pi} \, \mathbb{E}_{\mbpi} u(\mbx) + (1 - \bar{\alpha}) \, \max_{\mbpi \in \Pi} \, \mathbb{E}_{\mbpi} u(\mbx) \\
& = \bar{\alpha} \left[ \underline{a} \, u(x_1) + \frac{1}{3} \, u(x_2) + \left( \frac{2}{3} - \underline{a} \right) u(x_3) \right] + (1 - \bar{\alpha}) \left[ \left( \frac{2}{3} - \underline{a} \right) u(x_1) + \frac{1}{3} \, u(x_2) + \underline{a} \, u(x_3)\right] \\
& = \left[ \bar{\alpha} \underline{a} + (1 - \bar{\alpha}) \left( \frac{2}{3} - \underline{a} \right) \right] u(x_1) + \frac{1}{3} \, u(x_2)+ \left[ \bar{\alpha} \left( \frac{2}{3} - \underline{a} \right) + (1 - \bar{\alpha})\underline{a} \right] u(x_3).
\end{align*}
Given that $x_1\geq x_3$, we obtain
\begin{equation} \label{alpha}
\alpha = \frac{3}{2} \left [\bar{\alpha} \left( \frac{2}{3} - \underline{a} \right) + (1 - \bar{\alpha}) \underline{a} \right]. \tag{$\star$}
\end{equation}
Note that the coefficients on $u(x_1)$ and $u(x_3)$ are strictly positive and sum to $\tfrac{2}{3}$, as required.
\end{proof}

\begin{claim}
If the underlying model is MEU, then $\alpha \geq \tfrac{1}{2}$ in the kinked specification. If $\alpha \geq \tfrac{1}{2}$ in the kinked specification, then it can be explained as arising from the MEU model.
\end{claim}

\begin{proof}
In the MEU model, $\bar{\alpha} = 1$ and so, by (\ref{alpha}),
\[
\alpha = 1 - \frac{3}{2} \, \underline{a} \geq 1 - \frac{3}{2} \, \frac{1}{3} =\frac{1}{2}.
\]
Similarly, if we set $\bar{\alpha} = 1$, then
\[
\underline{a} = \frac{2}{3} \, (1 - \alpha).
\]
So long as $\alpha \geq \tfrac{1}{2}$, we obtain $\underline{a} \leq \tfrac{1}{3}$, as required.
\end{proof}

\begin{claim}
Ambiguity aversion is increasing in $\alpha$ and the decision maker is ambiguity seeking if $\alpha < \tfrac{1}{2}$ and ambiguity averse if $\alpha > \tfrac{1}{2}$.
\end{claim}

We must first be clear about what we mean when we say that a decision maker is ``ambiguity seeking,'' ``ambiguity averse,'' or ``more ambiguity averse'' than another. An \emph{Anscombe-Aumann act} specifies an objective lottery in each state. Let $\mathbf{f} = (f_1, f_2, f_3)$ denote an Anscombe-Aumann act, where $f_s$ is the objective lottery (over money) that occurs in state $s$, for $s = 1, 2, 3$. For any $t \in [0, 1]$, we denote by $t \mathbf{f} + (1 - t) \mathbf{g}$ the act where the lottery in state $s$ is $t f_s + (1 - t) g_s$. A decision maker with utility function $U$ is \emph{ambiguity averse} if, for any two acts $\mathbf{f}$ and $\mathbf{g}$ such that $U(\mathbf{g}) = U(\mathbf{f})$ then $U(t \mathbf{f} + (1 - t) \mathbf{g}) \geq U(\mathbf{f})$; the decision maker is \emph{ambiguity seeking} if the latter inequality is reversed (see \cite{gilboa1989}). A constant act $\mathbf{f}$ is one where $f_s = f_{s'}$ at any two states $s$ and $s'$; in other words, there is no uncertainty across states. The utility function $\tilde{U}$ is \emph{more ambiguity averse} than another utility function $U$ if, whenever $U(\mathbf{g}) \geq U(\mathbf{f})$ where $\mathbf{g}$ is a constant act, then $\tilde{U}(\mathbf{g}) \geq \tilde{U}(\mathbf{f})$.

\begin{proof}
For the act $\mathbf{f}$, the $\alpha$-MEU decision maker has expected utility $u^{f_s}$ for $f_s$. The utility of the act $\mathbf{f}$ is then
\[
U(\mathbf{f}) = \frac{2}{3} \, \alpha \min \, \{u^{f_1}, u^{f_3}\} + \frac{1}{3} \, u^{f_2} + \frac{2}{3} \, (1 - \alpha ) \max \, \{u^{f_1}, u^{f_3}\}.
\]
Similarly, let $\mathbf{g}$ be another Anscombe-Aumann act, with utility
\[
U(\mathbf{g}) = \frac{2}{3} \, \alpha \min \, \{u^{g_1}, u^{g_3}\} + \frac{1}{3} \, u^{g_2} + \frac{2}{3} \, (1 - \alpha) \max \, \{u^{g_1}, u^{g_3}\}.
\]
Notice that $U(\mathbf{f})$ is decreasing in $\alpha$ and $U(\mathbf{g})$ is constant if $g$ is a constant act. Therefore, if $U$ and $\tilde{U}$ correspond to lower and higher values of $\alpha$, then it is clear that a higher value of $\alpha$ implies greater ambiguity aversion.

For any $t \in [0, 1]$,
\begin{align*}
U(t \mathbf{f} + (1 - t) \mathbf{g}) = {}& \frac{2}{3} \, \alpha \min \, \{t u^{f_1} + (1 - t) u^{g_1}, t u^{f_3} + (1 - t) u^{g_3}\} \\
& + \frac{1}{3} \, (t u^{f_2} + (1 - t) u^{g_2}) \\
& + \frac{2}{3} \, (1 - \alpha) \max \, \{t u^{f_1} + (1 - t) u^{g_1}, t u^{f_3} + (1 - t) u^{g_3}\}.
\end{align*}
Ambiguity aversion requires that if $U(\mathbf{g}) = U(\mathbf{f})$ then $U(t \mathbf{f} + (1 - t) \mathbf{g}) \geq U(\mathbf{f})$, which holds if the function $F : \mbbR^3 \to \mbbR$ given by
\[
F(z_1, z_2, z_3) = \frac{2}{3} \, \alpha \min \, \{z_1, z_3\} + \frac{1}{3} \, z_2 + \frac{2}{3} \, (1 - \alpha) \max \, \{z_1, z_3\}
\]
is a concave function. This can be re-written as
\begin{align*}
F(z_1, z_2, z_3) & = \frac{2}{3} \, \alpha \, \frac{z_1 + z_3 - \vert z_1 - z_3 \vert}{2} + \frac{1}{3} \, z_2 + \frac{2}{3} \, (1 - \alpha) \, \frac{z_1 + z_3 + \vert z_1 - z_3 \vert}{2} \\
& = \frac{1}{3}(z_1+z_2+z_3)+\frac{1}{3}(1-2\alpha)|z_1-z_3|.
\end{align*}
The first term is linear in $(z_1,z_2,z_3)$ and the second term, $|z_1-z_3|$, is convex in $(z_1,z_3)$.  Therefore, $F$ is concave if $1-2\alpha\leq 0$ and convex if $1-2\alpha\geq 0$.
\end{proof}

\section{Rationalizability}

In the \cite{ahn2014} experiment, subjects are presented with a sequence of portfolio choice problems. There are three states of nature denoted by $s = 1, 2, 3$, and for each state $s$, there is a security that pays one token (the experimental currency) in state $s$ and nothing in the other states. State 2 has an objectively known probability of $\pi_2 = \tfrac{1}{3}$, whereas the probabilities of states 1 and 3 are unknown. Subjects are only informed that the two uncertain states (together) occur with probability $\tfrac{2}{3}$. The experiment has 50 rounds. In each round, the subject is presented with a budget set $\cal{B}^i = \{\mbx \in \mbbR^3_+ : \mbp^i \cdot \mbx \leq 1\}$ and chooses an allocation $\mbx^i$. The dataset collected from a subject thus has the form $\cal{D} = \{(\mbp^i ,\mbx^i)\}_{i = 1}^{50}$.

\paragraph{Rationalizability, GARP, and WARP}

A dataset $\cal{D} = \{(\mbp^i, \mbx^i)\}_{i = 1}^{50}$ is \emph{rationalizable} by utility maximization if there is a utility function $U : \mbbR^3_+ \to \mbbR$ such that $U(\mbx^i) \geq U(\mbx)$ for all $\mbx \in \cal{B}^i$. \citeauthor{afriat1967}'s (\citeyear{afriat1967}) theorem tells us that a dataset $\cal{D}$ is rationalizable by a \emph{well-behaved} (continuous and increasing) utility function $U$ if and only if $\cal{D}$ obeys the \emph{generalized axiom of revealed preference} (GARP).

GARP is an intuitive no-cycling condition on a dataset $\cal{D}$, which is easy to check in empirical practice. Let $\cal{X} = \{\mbx^i\}_{i = 1}^{50}$. For $\mbx^i$ and $\mbx^j$ in $\cal{X}$, we say that $\mbx^i$ is \emph{directly revealed preferred} to $\mbx^j$, denoted $\mbx^i \, R^D \, \mbx^j$, if $\mbp^i \cdot \mbx^j \leq 1$ (or equivalently, $\mbx^j \in \cal{B}^i$) and \emph{directly revealed strictly preferred} if the inequality is strict. The \emph{revealed preference} relation $R$ is the transitive closure of the direct revealed preference relation $R^D$: $\mbx^i \, R \, \mbx^j$ if there is a sequence $\{\mbx^k\}_{k = 1}^K$ with $\mbx^1 = \mbx^i$ and $\mbx^K = \mbx^j$, such that $\mbx^k \, R^D \,  \mbx^{k + 1}$ for every $k =  1, \ldots, K - 1$. GARP requires that if $\mbx^i \, R \, \mbx^j$ then $\mbx^j$ is not directly revealed strictly preferred to $\mbx^i$. In other words, GARP is violated whenever $\cal{D}$ contains a \emph{revealed preference cycle}: a sequence $\{\mbx^k\}_{k = 1}^K$ with $\mbx^1 = \mbx^i$ and $\mbx^K = \mbx^i$, such that $\mbx^k \, R^D \, \mbx^{k + 1}$ for every $k = 1, \ldots, K - 1$, where at least one direct revealed preference relation is strict; such a revealed preference cycle is referred to as a \emph{GARP violation}.

The \emph{weak axiom of revealed preference} (WARP) is a weakening of GARP which only requires that if $\mbx^i \, R^D \, \mbx^j$ then $\mbx^j$ is not directly revealed strictly preferred to $\mbx^i$. WARP is violated whenever $\cal{D}$ contains a pairwise revealed preference cycle: $\mbx^i \, R^D \, \mbx^j$ and $\mbx^j \, R^D \, \mbx^i$, where at least one direct revealed preference relation is strict; such a cycle is referred to as a \emph{WARP violation}. It is straightforward to show that WARP holds if demand is generated by a complete, but not necessarily transitive, preference ordering (see Appendix I in \cite{dembo2026}). When there are just two commodities, it is known that WARP and GARP are equivalent properties; in other words, a dataset obeys GARP if and only if it obeys WARP (see \cite{rose1958}). However, WARP is strictly weaker than GARP whenever there are three commodities or more (as in the \cite{ahn2014} experiment).

\paragraph{Rationalizability Index $\estar$}

\cite{afriat1973} provides an extension to \citeauthor{afriat1967}'s (\citeyear{afriat1967}) theorem which facilitates the calculation of the \emph{critical cost-efficiency index} (CCEI). We say that \emph{$\cal{D} = \{(\mbp^i, \mbx^i)\}_{i = 1}^{50}$ is rationalizable at cost-efficiency $e \in (0, 1]$} if there is a well-behaved utility function $U : \mbbR^3_+ \to \mbbR$ such that
\begin{equation} \label{ccei}
U(\mbx^i) \geq U(\mbx) \; \; \text{for all} \; \; \mbx \in \cal{B}^i(e) = \{\mbx \in \mbbR^3_+ : \mbp^i \cdot \mbx \leq e\}. \tag{$\star\star$}
\end{equation} 
The CCEI (which we refer to as the rationalizability score $\estar$ in the main paper) is the supremum of $e$ such that $\cal{D}$ is rationalizable at cost-efficiency $e$. Whether or not a dataset $\cal{D}$ is rationalizable at cost-efficiency $e$ can be checked by a suitable extension of GARP.

Given $e \in (0,1]$, for $\mbx^i$ and $\mbx^j$ in $\cal{X}$, we say that $\mbx^i$ is \emph{directly revealed preferred at cost-efficiency $e$} to $\mbx^j$, denoted $\mbx^i \, R^D(e) \, \mbx^j$, if $\mbp^i \cdot \mbx^j \leq e$ (or equivalently, $\mbx^j \in \cal{B}^i(e)$); if this inequality is strict, we say that $\mbx^i$ is \emph{directly revealed strictly preferred at cost-efficiency $e$} to $\mbx^j$, denoted $\mbx^i \, P^D(e) \, \mbx^j$. We denote the transitive closure of $R^D(e)$ by $R(e)$.  We say that  $\mbx^i$ is revealed preferred at cost efficiency $e$ to $\mbx^j$ if $\mbx^i \, R(e) \, \mbx^j$. A dataset $\cal{D}$ satisfies \emph{GARP at cost-efficiency $e$} if, whenever $\mbx^i \, R(e) \, \mbx^j$, then it is not the case that $\mbx^j \, P^D(e) \, \mbx^i$. \cite{afriat1973} shows that \emph{$\cal{D}$ is rationalizable at cost-efficiency $e$ if and only if it obeys GARP at cost-efficiency $e$}. Checking whether $\cal{D}$ obeys GARP at a given cost-efficiency $e$ is computationally straightforward. The value of $\estar$ can then be obtained through a binary search.

\paragraph{Rationalizability Index $\edoublestar$}

A dataset $\cal{D}$ is \emph{FOSD-rationalizable under the uniform prior} if it is compatible with a well-behaved and symmetric utility function $U$. An extension of \citeauthor{afriat1967}'s (\citeyear{afriat1967}) theorem by \cite{nishimura2017} provides a GARP-like condition that characterizes those datasets that can be rationalized by a utility function that is increasing with respect to a given preorder. This result can be applied to check whether $\cal{D}$ is FOSD-rationalizable under the uniform prior and, more generally, whether $\cal{D}$ is FOSD-rationalizable under the uniform prior at a given cost-efficiency $e$ (whether there is a well-behaved and symmetric utility function $U$ such that the optimality condition (\ref{ccei}) holds).

For $\mbx^i$ and $\mbx^j$ in $\cal{X}$, we say that $\mbx^i$ is \emph{directly revealed FOSD-preferred (directly revealed strictly FOSD-preferred) under the uniform prior at cost-efficiency $e$} to $\mbx^j$ if there is some $\mby$, the entries of which are a permutation of the entries in $\mbx^j$, such that $\mbp^i \cdot \mby \leq (<) \, \, e$; we denote these relations by $\bar{R}^D(e)$ and $\bar{P}^D(e)$, and the transitive closure of $\bar{R}^D(e)$ by $\bar{R}(e)$. A dataset $\cal{D}$ obeys \emph{FOSD-GARP under the uniform prior at cost-efficiency $e$} if, whenever $\mbx^i \, \bar{R}(e) \, \mbx^j$, then it is not the case that $\mbx^j \, \bar{P}^D(e) \, \mbx^i$. We use this modified GARP property to check if $\cal{D}$ can be rationalized at cost-efficiency $e$ by a well-behaved and symmetric utility function; \cite{nishimura2017} guarantees that \emph{$\cal{D}$ is FOSD-rationalizable under the uniform prior at cost-efficiency $e$ if and only if it obeys FOSD-GARP at cost-efficiency $e$}. Checking whether $\cal{D}$ obeys FOSD-GARP at a given cost-efficiency $e$ is computationally straightforward. The value of $\edoublestar$ can then be obtained through a binary search.

\paragraph{Rationalizability Index $\tildeedoublestar$}

A dataset $\cal{D}$ is \emph{FOSD-rationalizable under multiple priors} if it is compatible with a well-behaved utility function $U$ that is ($i$) symmetric between the uncertain states 1 and 3, and ($ii$) satisfies  $U(b, a, b) < U(a, b, a)$ whenever $0 \leq b < a$. We can formulate a necessary and sufficient test for whether $\cal{D}$ is FOSD-rationalizable under multiple priors at a given cost-efficiency $e$ by appealing to \cite{nishimura2017}. A dataset $\cal{D}$ obeys \emph{FOSD-GARP under multiple priors at cost-efficiency $e$} if whenever $\mbx^i \, \widetilde{R}(e) \, \mbx^j$, then it is not the case that $\mbx^j \, \widetilde{P}^D(e) \, \mbx^i$. It is straightforward to check that this modified GARP property is necessary for $\cal D$ to be rationalized at cost-efficiency $e$ by a well-behaved utility function obeying properties ($i$) and ($ii$) above; the result of \cite{nishimura2017} guarantees that this property is also sufficient. The value of $\tildeedoublestar$ can then be obtained through a binary search.

\paragraph{Rationalizability Index $\etriplestar$}

A dataset $\cal{D}$ is \emph{EUT-rationalizable} if there is an increasing and continuous Bernoulli index $u : \mbbR_+ \to \mbbR$ such that $U(\mbx) = \tfrac{1}{3} u(x_1) + \tfrac{1}{3} u(x_2) + \tfrac{1}{3} u(x_3)$ rationalizes $\cal{D}$. If we can find a $U$ of this form such that the optimality condition (\ref{ccei}) holds, then we say that $\cal{D}$ is \emph{EUT-rationalizable at cost-efficiency $e$}. The necessary and sufficient conditions to check for EUT-rationalizability at a given cost-efficiency are developed in \cite{polisson2020}. This test is implemented in \cite{polisson2020} and also in \cite{dembo2026}, and the reader may consult those papers for the details. The value of $\etriplestar$ can then be obtained through a binary search.

\end{document}